\documentclass[12pt]{article}
\usepackage[export]{adjustbox}
\usepackage{authblk}
\usepackage{bbm}
\usepackage [autostyle, english = american]{csquotes}
\MakeOuterQuote{"}
\usepackage{pgfpages}
\usepackage{graphicx}
\usepackage{multicol}
\usepackage{csquotes}
\usepackage{multirow}
\usepackage{graphicx} 
\usepackage{booktabs} 
\usepackage{authblk}
\usepackage{caption}
\usepackage{subcaption}
\usepackage{stackengine}
\usepackage{amsmath}
\usepackage{comment}
\usepackage{hyperref}
\hypersetup{
    colorlinks=true,
    linkcolor=blue,
    filecolor=blue,      
    urlcolor=blue,
    citecolor=blue,
}
\usepackage[spanish,english]{babel}

\usepackage{amssymb}
\usepackage{stackrel}
\usepackage{natbib} 

\usepackage[capposition=top]{floatrow}

\usepackage{lscape}

\usepackage[margin=1in]{geometry}

\title{\Large Does an increase in the cost of imported inputs hurt exports? \\ \large Evidence from firms' network of foreign suppliers}

\author[1]{Santiago Camara}
\affil[1]{Northwestern University \& Red-NIE}
\date{\today}

\begin{document}

\maketitle

\begin{abstract}
    This paper examines the relationship between changes in the cost of imported inputs and export performance using a novel dataset from Argentina which identifies domestic firms' network of foreign suppliers. To guide my empirical strategy, I construct a heterogeneous firm model subject to quality choice and frictions in the market for foreign supplier. The model predicts that the impact of an increase in the cost of imported inputs to be increasing in the adjustments costs of supplier linkages and in the quality of the product exported. I take the model to the data by constructing firm-specific shocks using a shift-share analysis which exploits firms' lagged exposure to foreign suppliers and finely defined import price shifts. Evidence suggests the presence of significant adjustment cost in firms' foreign supplier linkages and strong complementarities between imported inputs and export performance, particularly of high-quality products.
    
    \medskip
    \medskip

\end{abstract}

\newpage
\section{Introduction} \label{sec:introduction}

\noindent
Emerging Market economies rely heavily in imported intermediate inputs and capital goods (see \cite{eaton2001trade} for example). In these countries, exporting firms show a particularly dependence of imports as they represents a significant share of their total production inputs (see \cite{kugler2009plants,brambilla2012exports}). The purchase of imported inputs implies establishing and maintaining firm-to-firm linkages with foreign firms. The inherent frictions pervasive to the market for input suppliers in domestic markets (see \cite{giulietti2014estimation} for energy suppliers and \cite{mortensen1999new} for labor supply) are exacerbated in international markets (see \cite{startz2016value}). Not only the quality of firm-to-firm matches is slowly revealed, but repeated interactions with suppliers usually involve significant informational frictions (see \cite{macleod2007reputations} and \cite{antras2015poultry}). The presence of these frictions imply that linkages with foreign suppliers are costly to adjust and not quickly substitutable (see \cite{huneeus2018production}). This paper seeks to quantify the adjustment costs of firms' foreign linkages' and the degree of complementarities between foreign inputs and export performance.

The impact of an increase in the cost of imported inputs on export performance is a classic question in the international trade literature. \cite{kasahara2013productivity} showed that more productive firms tend to import and export more, and argued that due to import and export complementarities, increases in the cost of foreign intermediate inputs can have a large adverse effect on the exports of final goods. The literature on international trade has also emphasized the importance of imported inputs for the production of high-quality goods (see \cite{fan2015trade} and \cite{bas2015input}). \cite{bastos2018export} show that firms which source higher-quality imported inputs tend to export higher quality products to richer and more selective countries. 

Studying the complementarity between imported inputs and exports is of particular importance for Emerging Market economies, like Argentina. These economies generally lack sophisticated domestic network of high-quality intermediate inputs (see \cite{kugler2009plants}) and usually have export bundles which rely heavily in low-value-added commodity goods (see \cite{statecommodityUNCTAD}). Thus, to achieve high levels of income, it is crucial for firms in these countries to access imported inputs to produce and export high-quality and high-value-added goods (see \cite{de1997export}) and \cite{gonzalez2012insertion}). 

To quantify the complementarities between imported inputs and export performance and the adjustment costs of firms' linkages with foreign suppliers I use a unique data set of Argentinean firms which identifies both the domestic importing firm and the foreign exporting supplier.\footnote{Appendix \ref{subsec:appendix_additional_evidence} presents evidence on the aggregate relevance of imports of inputs and capital. In that Appendix, Figure \ref{fig:impo_bec} shows that imports of intermediate and capital goods explain between 80\% and 90\% of Argentina's total imported value for the period 1994-2019. Figure \ref{fig:impo_investment} shows that imported capital goods explain between 55\% and 65\% of the total investment in machinery and/or equipment for the same period.} This dataset allows me to observe domestic firms' network of foreign suppliers and their imported-input sourcing strategy. In particular, I can observe trade flows at the ``domestic firm''-``foreign supplier''-``product''-''source country'' level. This allows me to observe finely defined changes in firms' cost of imported inputs.

To organize my empirical work, I construct a heterogeneous firm model subject to quality choice which can sell to foreign markets with different preferences for quality. Domestic firms are heterogeneous both in their ability to produce using lower amounts of inputs and in their ability to produce high-quality products at low costs (see \cite{hallak2013product}). Crucially, domestic firms face search and matching frictions in the market for foreign suppliers which differ in their efficiency to supply inputs at a low cost. Although firms have incentives to match with the most efficient supplier, searching involves the payment of a fixed cost and matching is random. The model's features produce a framework in which firms' export scope and performance is appreciably heterogeneous but still delivers two sharp predictions. First, the impact of an imported cost shock is increasing in the severity of search and matching frictions in the market for foreign suppliers. Second, the impact of an imported cost shock on exports is increasing in the role of imported in production and in the destination market's preference for quality. 

While the prediction that an increase in imported inputs negatively affects export performance is straightforward, taking it to the data presents a challenge. First, firms endogenously choose their network of foreign suppliers (see instance \cite{dhyne2021trade}). This implies that firms observe market prices and endogenously choose from whom to buy and how much. To properly deal with this empirical challenge I use a shift-share analysis as presented by \cite{ad2019shift}. I measure a firm's exposure to import cost shocks using its lagged shares of imported value at the ``foreign-supplier''-``product''-``source-country'' level. I measure price shifts as the change in the average price a foreign-supplier charges for a product coming from a specific source country to all Argentinean firms. I find that a one standard deviation import cost shock leads to a $7.44\%$ drop in imported quantities and a persistent $2.44\%$ drop in exported quantities up to 2 years after the initial shock. These results suggest that the presence of costly to adjust frictions between Argentinean firms and their foreign suppliers, and strong complementarity between imported inputs and exported quantities. 

Finally, while \textit{a priori} one would expect high-quality products to be impacted harder by a rise in imported input costs, this is also a challenging prediction to test in the data. Direct measures of product quality are not available in the Argentinean customs data. To test the differential impact of an import cost shock across products I estimate my empirical model across different sample partitions. I proxy product quality exploiting destination market's level of income and by exploiting export's degree of product differentiation. Through these sample partition exercises I find that the complementarities between imported inputs and export performance is twice as big for high-quality goods compared to the full sample results.

\section{Theoretical Framework} \label{sec:theoretical_framework}

This section of the paper builds a model of heterogeneous firms subject to quality choice and search and matching frictions in the market for foreign suppliers. The model is developed in partial equilibrium and time is discrete, indexed by $t = 0,1,2,\ldots$ . I assume that every period firm can only be matched with one foreign supplier.\footnote{While this may seem as a far-fetched assumption, recent evidence suggests that domestic firms' network of foreign suppliers is granular or limited, see \cite{bernard2018firm}.} Firms' decision problem can be divided into two sub-periods: a search and matching stage and a production and sale stage. 

Firm $i$ maximizes the discounted present value of profits
\begin{align*}
    \Pi = \sum^{\infty}_{t=0} \beta \pi \left(z_i,\xi_i,c_{j,t}\right)
\end{align*}
where $\{x_i,\xi_i \}$ are firms attributes described below and $c_{j,t}$ represents the efficiency of supplier $j$ with which firm $i$ is matched with in period $t$. The timing of the model is the following: (i) period $t$ starts with firm $i$ observing the efficiency of supplier $j$ she is currently matched with, (ii) firm decides whether to search or not, and if the firm searches, whether to remain with the current supplier or change to the newly matched supplier, (iii) finally, firm $i$ decides which markets to serve, produces and sells.

\noindent
\textbf{Production \& sale stage.} I build on \cite{hallak2013product} and \cite{bastos2018export} and assume that firms differ across two dimensions: (i) "process productivity" $z_i$ which denotes firm $i$'s ability to produce with a lower amount of inputs; (ii) “product productivity” $\xi_i$ as the ability to produce quality incurring lower fixed outlays. Both $z_i$ and $\xi_i$ are exogenously distributed across domestic firms. 

In each country, there is a representative consumer which demands goods according to the following utility function
\begin{equation*} \label{eq:utility_function_quality_one}
    U_d = \left[\sum_{i \in I_d} \left(\lambda^{\zeta \left(y_d\right)}_{i,d}  x_{i,d} \right)^{\frac{\rho-1}{\rho}} \right]^{\frac{\rho}{\rho-1}}
\end{equation*}
where $I_d$ is the set of firms selling in country $d$; $\lambda_{i,d}$ is product quality; $x_{i,d}$ is the quantity consumed and; $\zeta_{d} \left(y_d\right)$ captures the representative consumer's valuation of quality. To simplify notation, I write $\zeta_d = \zeta \left(y_d\right)$. This preferences give way to the following demand function for firm $i$'s output coming from country $d$ as
\begin{equation} \label{eq:consumer_demand_quality_one}
    x_{i,d} = \frac{\left(p_{i,d}\right)^{-\rho} \left(\lambda_{i,d}\right)^{\left(\rho-1\right) \zeta_d} y_d }{P^{-\rho}_d}
\end{equation}
where $p_{i,d}$ is the price charged by firm $i$ in country $d$, and $P_d$ is the country-wide quality-adjusted price index:
\begin{equation*} \label{eq:price_index_quality_one}
    P_d = \left[\sum_{i\in I_d} \left(\frac{p_{i,d}}{\lambda^{\zeta_d}_{i,d}}\right)^{1-\rho} \right]^{\frac{1}{1-\rho}}
\end{equation*}
Note that demand is increasing in product's quality, governed by parameter $\zeta_d$.

I assume that firms use a constant return to scale technology on separate production lines for each of the $d$ markets served.\footnote{This assumption is common in the literature of international trade (see for instance \cite{bastos2018export}).} In particular, the production technology is a Cobb Douglas production function. 
\begin{align} 
    x_{i,d} &= z_i L^{\alpha}_{i,d} M^{1-\alpha}_{i,d}  \label{eq:production_function_tier1_one}
\end{align}
where $L_{i,d}$ and $M_{i,d}$ represent firm $i$'s labor and imported input quantities. Firm $i$'s marginal cost on production line $d$ can be expressed as a function of wages, $w$, and the matched supplier's efficiency levels $c_j$.
\begin{align} 
    C_{i} \left(z_i, c_{j} \right) &= \frac{1}{z_{i}} \frac{1}{\alpha^{\alpha}(1-\alpha)^{1-\alpha}} w^{\alpha} \left( c_j\right)^{1-\alpha} \label{eq:marginal_cost}
\end{align}

I assume that product quality also involves incurring costs. I introduce a quality cost specification similar to that presented by \cite{hallak2013product}. Quality costs are represented by the following function
\begin{equation}
    F^{q}\left(\xi_i \right) =  \frac{f}{\xi_i} \lambda_{i,d}
\end{equation}
where $f$ is a constant and $\lambda_{i,d}$ represents firm $i$'s quality choice. Note that firms with higher $\xi_i$ are able to produce high-quality products at lower costs. These costs can be thought of as product design and development costs. Lastly, in order to export to destination market $d$, a firm must pay a market specific fixed cost $F^{d}$.

\noindent
\textbf{Profit Maximization.} Firm $i$'s production line $d$'s profit maximization problem can be stated as
\begin{equation} \label{eq:profit_maximization_text}
\begin{aligned}
\max_{\lambda_{i,d},x_{i,d},p_{i,d}} \pi_{i,d} \left(z_i, \xi_i, c_{j} \right) = \quad & \left(p_{i,d} - C_{i} \right) x_{i,d} - \frac{f}{\xi_i}\lambda_{i,d} - \mathbbm{1}\left[x_{i,d}>0\right] \times F^d \\
\textrm{s.t.} \quad & x_{i,d} = \frac{\left(p_{i,d}\right)^{-\rho} \left(\lambda_{i,d}\right)^{\left(\rho-1\right) \zeta_d} y_d }{P^{-\rho}_d} \\
  & C_{i} \left(z_i, c_{j} \right) = \frac{1}{z_{i}} \frac{1}{\alpha^{\alpha}(1-\alpha)^{1-\alpha}} w^{\alpha} \left( c_j\right)^{1-\alpha} 
\end{aligned}
\end{equation}
This framework allows for rich a heterogeneity in firms' export performance and scope. Firms' differences in attributes $\{x_i,\xi_i\}$ imply that differences in both quantities sold in each market, but also differences in export scope. For instance, firms with high $x_i$ but low $\xi_i$ might be able to serve low-income countries but may not be able to serve high-income countries due to the high costs they face in order to produce quality goods.\footnote{Appendix \ref{subsubsec:appendix_model_features} provides additional details and characterizes firms' export scope through a calibrated numerical exercise}. 

Firm $i$'s optimal quality and quantities sold in destination market $d$ can be expressed as
\begin{align}
    \lambda^{*}_{i,d} &= \Lambda_1 \times C^{\frac{\rho-1}{\left(\rho-1\right)\zeta_d-1}}_{i,d} \label{eq:optimal_quality_short} \\
    x^{*}_{i,d} &= \Lambda_2 \times  C^{\frac{\rho-1}{\left(\rho-1\right)\zeta_d-1}-1}_{i,d} \label{eq:optimal_quantity_short}
\end{align}
where $\Lambda_1$ and $\Lambda_2$ are constant which depend on model parameters and exogenous variables.\footnote{See Appendix for more model details \ref{subsec:appendix_model_details}.} Note that as long as condition
\begin{align} \label{eq:condition_profit_KEY}
    \zeta_d < \frac{\rho}{\rho-1}
\end{align}
is satisfied, the optimal quality and optimal quantities sold are decreasing in $C_i$. This condition implies that consumers' love for quality, represented by $\zeta_d$, can not outweigh the demand's price elasticity of substitution. If the condition is not met, the problem boils down to a rather uninteresting case in which firms' optimal behavior is to rise quality and prices to infinity.\footnote{This condition is also necessary for the second order conditions to be met, see Appendix \ref{subsubsec:appendix_profit_maximization_SOC}.}

Next, I characterize how an increase in the cost of imported input $c_j$ impacts quantities sold in market $d$. By taking logarithms on both sides of Equation \ref{eq:optimal_quantity_short} and using the expression from $C_i$ in Equation \ref{eq:marginal_cost}, the impact of an increase in $c_j$ on $x^{*}_{i,d}$ is given by
\begin{align}
    \frac{\partial \ln x^{*}_{i,d}}{\partial \ln c_{j}} = \left( \frac{\rho-1}{\left(\rho-1\right)\zeta_d-1}-1 \right) \left(1-\alpha\right)
\end{align}
Note that this impact is increasing in the importance of imported inputs in production, $1-\alpha$, and in destination country consumer's love for quality, $\zeta_d$. To understand the intuition behind love for quality amplifying the impact of changes in marginal costs on quantities sold it is useful to study firms' first order conditions. Re-writing firm $i$'s profit maximization in Equation \ref{eq:profit_maximization_text} in terms of optimal quantities and quality leads to the following two first order conditions
\begin{align}
    \frac{\partial \pi }{\partial x_{i,d}}: & \quad \psi x^{-\frac{1}{\rho}}_{i,d} \left(\lambda_{i.d}\right)^{\frac{\left(\rho-1\right)\zeta_d}{\rho}} = C_{i} \left(z_i, c_j\right)  \label{eq:FOC_quantities_text} \\
    \frac{\partial \pi }{\partial \lambda_{i,d}}: & \quad \psi x^{1-\frac{1}{\rho}}_{i,d} \left(\lambda_d\right)^{\frac{\left(\rho-1\right)\zeta_d}{\rho}-1}  \zeta_d = \frac{f}{\xi_i} \label{eq:FOC_quality_text}
\end{align}
where $\psi = (\rho-1)/\rho \times y^{\frac{1}{\rho}}_d P_d$. Equation \ref{eq:FOC_quantities_text} shows that an increase in $C_{i} \left(z_i, c_j\right)$ leads to a reduction in quantities sold. Simultaneously, the left hand side of Equation \ref{eq:FOC_quality_text} shows that the marginal benefit of quality is increasing in quantities sold $x_{i,d}$ and $\zeta_d$. Consequently, an increase in $C_{i}$ which reduces optimal quantities $x_{i,d}$ leads to a larger drop in the marginal benefit of quality the greater $\zeta_d$ is. In turn, this leads to a drop in firms' optimal quality choice which reinforces the drop in optimal quantities sold due to lower consumer demand. Thus, an increase in the marginal cost $C_i$ triggers a reduction in optimal quantities and qualities which reinforce each other, leading to an amplification of the initial shock, which is increasing in consumer's love for quality. 

\noindent
\textbf{Search \& matching stage.} I build on \cite{defever2017supplier} and assume that there are many foreign suppliers that can potentially match with domestic firms and produce the imported input. Foreign suppliers differ in their efficiency or marginal cost $c$ of producing the input. Marginal costs are distributed across potential suppliers following a probability density function $g(c)$ with support $\left[\underline{c},\bar{c} \right]$, where $\underline{c}>0$. 

Under this framework, a domestic firm has an incentive to match with a supplier with the lowest possible marginal cost $c$. In order to search for a new supplier, domestic firms must pay a fixed cost $F$.\footnote{Note that I abstract from any potential strategic behavior and assume that foreign suppliers have a passive role} Firms' search is not direct, i.e., can not choose with which foreign supplier to match. Finally, I assume that after searching and observing the foreign supplier's efficiency, domestic firms have the option to remain matched with her old supplier.

I turn to characterizing the conditions under which a domestic firm searches for a new supplier. Firm $i$ searches for a new foreign supplier if the expected discounted gains of matching with a more efficient supplier outweigh the fixed cost of searching. Denoting firms' current supplier's efficiency by $\tilde{c}$, and given that the search problem is the same every period, the search condition boils down to
\begin{equation} \label{eq:IC_search_prop1}
    \frac{ \{ \mathbb{E} \left[ \pi \left(z_i,\xi_i,\tilde{c}\right) | c<\tilde{c} \right] - \pi \left(z_i,\xi_i,\tilde{c}\right) \} G\left(\tilde{c}\right) }{1-\beta}  \geq F
\end{equation}
The term $\{ \mathbb{E} \left[ \pi \left(z_i,\xi_i,\tilde{c}\right) | c<\tilde{c} \right] - \pi \left(z_i,\xi_i,\tilde{c}\right)$ on the left hand side's numerator reflects the expected profit differential if the searching firm matches with a supplier with higher efficiency, i.e. $c < \tilde{c}$. In this case, the firm changes foreign suppliers and experiences an increase in profits. The second term on the numerator, $G\left(\tilde{c}\right)$ represents the probability of matching with a foreign supplier more efficient than $\tilde{c}$. Given that if the firm matches with a more efficient supplier, this becomes her default benchmark for period $t$ onward, the firm discounts the gains of searching by $1-\beta$.

The condition in Equation \ref{eq:IC_search_prop1} implies that the higher the fixed costs $F$, the less likely it is for a firm to search for new suppliers. Also, the less efficient firm $i$'s current supplier is, the greater the expected profit differential of matching with a new supplier. Furthermore, the less efficient firm $i$'s current supplier is, the higher the probability of matching with a more successful supplier, i.e. higher $G\left(\tilde{c}\right)$, increasing firm $i$'s incentive to search. 

\noindent
\textbf{Model predictions.} The model constructed above yields sharp predictions over the impact of an increase in $c_j$ on firms' export performance. First, it is straightforward to see that the magnitude and persistence of the impact depend positively on the adjustment cost of domestic firms' supplier linkages. This is represented by the model's fixed cost of searching $F$ and the random match with heterogeneous suppliers, represented by $G(c)$. If fixed cost $F$ is zero and firms can substitute toward more efficient suppliers immediately, an increase in $c_j$ should lead to no impact on firms' export performance. Consequently, the higher the fixed cost $F$ and the lower the probability of matching with a more efficient supplier, the greater and more persistent the impact of the shock will be. 

Second, given the search and matching frictions in the market for foreign suppliers which determine firms' supplier efficiency, the model provides sharp predictions on export performance. A rise in $c_j$ increases firms' overall marginal cost $C_i$ and thus negatively impacting sales. The impact magnitude depends positively on the relative importance of imported inputs in production, $1-\alpha$, and on the destination market's love for quality $\zeta_d$. While my dataset does not allow me to estimate firm's production functions and import intensity, it does provide with proxy variables for product quality. This allows me to test the prediction that exports of higher-quality goods are more sensitive to import cost shocks.

\section{Data \& Identification Strategy} \label{sec:data_empirical_methodology}

In this section of the paper I describe the datasets used in this paper and the identification strategy proposed to identify import cost shocks.

\noindent
\textbf{Data description.} I use data from administrative customs records, based on customs declarations forms collected by Argentina's customs office, which comprises the entire universe of Argentinian trade flows for the period 2000-2008. Information is dis-aggregated by firm, date of shipment, product at 6-digit HS, exports by destination and imports by source country. For import flows, I am able to identify the foreign supplier firm. Since foreign suppliers do not have a unique ID that identifies them (I only observe the firms' name declared in customs' form), their registration is not harmonized and is subject to errors. Therefore, I carry out a cleaning procedure that allows me to join the information of suppliers that were registered in two or more different ways. Details of this cleaning procedure and secondary datasets used can be found in Appendix \ref{subsec:Appendix_cleaning} and Appendix \ref{subsec:appendix_additional_data}, respectively. 

\noindent
\textbf{Identification strategy.} Based on firms' search and matching frictions described in Section \ref{sec:theoretical_framework}, I follow \cite{hummels2014wage}, \cite{huneeus2018production} and \cite{bernard2020heterogeneous} and use a shift share design to construct firm-level shocks. I define an import cost shock for firm $i$ in year $t$, as
\begin{equation} \label{eq:shift_share}
    \Delta \log p^{I}_{i,t} = \sum_{k \in \Omega^{I}_{i,0}} s^I_{ik,0} \Delta \log p_{k,t}
\end{equation}
where $\Omega^{I}_{i,0}$ is the set of foreign suppliers $k$ that firm $i$ sources from in period $t=0$, $s^I_{ik,0}$ is the share of imports of firm $i$ from supplier $k$ in $t=0$ and $\Delta \log p_{k,t}$ is the percentage growth of the import price from supplier $k$. Before defining in detail what a foreign supplier $k$ means in my setting, I address two issues: (i) discuss the usefulness and appropriateness of the shift-share analysis,  (ii) which measure of price change to use for $\Delta \log p_{k,t}$.

To begin with, I momentarily assume that the measure of changes in import prices is exogenous and discuss how a shift-share analysis is appropriate for the current analysis. Even if import prices are exogenous to Argentinean firms, firms choose their network of suppliers, composition of their input bundle and source countries endogenously. I use \cite{ad2019shift}'s shift-share design which assumes the exogeneity of the shifts rather than from the shares. The reasoning behind this choice is that firms choose which suppliers $k$ to trade with and are subject to the shocks generated through these established linkages. When firms make the decision about which foreign supplier $k$ to trade with, future shocks might be already in firms' informational set. In order to address any bias arising from anticipation behavior I used lagged shares of exposure (in particular, I construct the shocks using period $t-1$, i.e., year $t-1$ weights). Note that this is in line with the search and matching stage of firms described in Section \ref{sec:theoretical_framework}, where firms at the start of period $t$ observe their current supplier's efficiency and can search in international markets before producing.

The shift-share analysis provides a source of significant heterogeneity in the cost shock across firms. This significant heterogeneity comes from the high dimensional space of foreign suppliers the Argentinean firms can choose from. For instance, Argentinean firms imported from 220 different countries, 4,961 different products and from 512,915 different suppliers during the period 2000-2008. This high degree of heterogeneity in sourcing strategies makes it unlikely that any two Argentinean firms are exposed exactly the same to the same set of foreign suppliers. A second source of heterogeneity across firms is each firms' link intensity or amount traded in said trade flow. 

Next, I turn to discussing the appropriate price change to be considered as shifts in Equation \ref{eq:shift_share}. An initial candidate as a shift would be to use log import price changes at the domestic firm $i$, foreign supplier $s$, product $p$ and destination country $d$
\begin{align*}
    \Delta \log p_{i,s,p,d,t} =  \log p_{i,s,p,d,t} - \log p_{i,s,p,d,t-1}
\end{align*}
which implies defining $k$ as $k = \{i,s,p,d\}$. For this specification to be a valid representation of the framework presented in Section \ref{sec:theoretical_framework} two assumptions must be met. First, firm $i$ has no bargaining power on the determination of its import prices, i.e. take foreign supplier $s$ price as exogenous. Second, for this shift to represent a true exogenous increase on its import prices firm $i$ must have no firm-specific shock which increases the cost of its inputs across every possible combination of foreign-supplier, product and source country.\footnote{For instance, a shock that increases firm $i$'s cost of international trade, such as distribution costs.} Additionally, this specification of price shift is highly susceptible to a attrition. This is, if firm $i$ does not buy in period $t$ from the same supplier, the same product from the same source country my dataset will not report price $p_{i,s,p,d,t}$, and I will impute a price-shift $\Delta \log p_{i,s,p,d,t} = 0$.

Given the concerns described above I define the price shift in Equation \ref{eq:shift_share} at the foreign supplier $s$, product $p$ and destination country $d$. This is, I define supplier $k$ as $k=\{s,p,d\}$
\begin{align*}
    \Delta \log p_{s,p,d,t} =  \log p_{s,p,d,t} - \log p_{s,p,d,t-1}
\end{align*}
Under this empirical specification, I assume that firm $i$ takes the average price supplier $s$ charges for product $p$ from destination country $d$ to all Argentinean firms. This is a less extreme assumption as it does not rely on assumptions on firm $i$'s direct behavior and is less prone to attrition.\footnote{The latter is due to it being more likely that a specific link between domestic $i$ and foreign supplier $s$ breaks than foreign supplier $s$ breaking all of her linkages with Argentinean firms.} Appendix \ref{subsec:appendix_additional_shocks} presents basic descriptive statistics of this shock specification.\footnote{Appendix \ref{subsec:appendix_additional_shocks} also presents three other shock specifications in which prices are defined at the: (i) ``firm $i$''-``supplier''-``product''-``source'' level as defined above, (ii) at the ``all firms except firm $i$''-``supplier''-``product''-``source'' level; and at the ``product''-``source'' level. This last specification is a standard specification in the literature, used for instance by \cite{huneeus2018production}. Results are robust across specifications which exploit the supplier dimension of the data set, but not the specification which computes price shifts at the ``product''-``source'' level.}

\section{Results} \label{sec:results}

This section presents the main results of the paper. I start by estimating the impact of an import cost shock on firms' imported quantities, which allows me to quantify adjustment costs on domestic firms' linkages with foreign suppliers. Then, I estimate the impact of an import cost shock on firms' exported quantities in order to quantify the complementarities between imported inputs and exports. Finally, I test whether high-quality exports are more sensitive to changes in the price of imported inputs. 

\noindent
\textbf{Costly to adjust input linkages.} I start by estimating the impact of an import cost shock on firms' imported quantities. To do so, I estimate the following  regression
\begin{align} \label{eq:intensive_imports}
    \ln QI_{i,p,d,t} &= \alpha_0 + \gamma_{i,p,d} +\gamma_t+ \beta^{M} \text{Shock}_{i,t} + \Gamma_{i,p,d,t} + \epsilon_{i,p,d,t}
\end{align}
where $\ln QI_{i,p,d,t+j}$ represent the log total quantities imported by firm $i$ of product $p$, from source country $d$ in year $t+j$ for $j\in\{0,1\}$, $\gamma_{i,p,d}$ is a firm - product- source country fixed effect, $\gamma_t$ is a time fixed effect, $\text{Shock}_{i,t}$ is the shock constructed in Section \ref{sec:data_empirical_methodology}, and $\Gamma_{i,p,d,t}$ is a set of controls which include lagged values of $\text{Shock}_{i,t}$ and $\ln QI_{i,t}$. Parameter $\beta^{M}$ represents imported quantities semi-elasticity to an import cost shock. The standard errors of the estimated parameters are computed using two-way clustering, at the firm $i$ and source country $d$ level.\footnote{The reasoning behind this clustering strategy is twofold. First, the constructed exogenous shocks are defined at the firm level, thus it implies clustering at the treatment level unit. Second, I want to incorporate source country level information.} 

Table \ref{tab:import_cluster} presents estimated values of $\beta^M$ on impact at time $t$, and for the year after the shock, time period $t+1$.
\begin{table}[ht]
    \centering
    \caption{Cost Shocks and Imported Quantities 2001 - 2008}
    \label{tab:import_cluster}
    \small
\begin{tabular}{lcc} 
  & \multicolumn{2}{c}{Log - Quantities Imported - $\ln Q_{i,p,d,t+j}$} \\
  & (1) & (2) \\
  & Period $t$ & Period $t+1$ \\ \hline \hline
$\beta^M-\text{Shock}_{i,t}$ & -0.305*** & -0.0465 \\
 & (0.0498) & (0.0579) \\
 &  &   \\
 Observations & 354,538 & 234,709 \\ \hline \hline
\multicolumn{3}{c}{\footnotesize Two Way Clustered (Firm - Source Country) Standard Errors in Parentheses} \\
\multicolumn{3}{c}{\footnotesize *** p$<$0.01, ** p$<$0.05, * p$<$0.1} \\
\end{tabular}
\floatfoot{\footnotesize \textbf{Note:} The regressions are computed controlling for the first and second lagged values of the dependent variable $\ln QI_{i,p,d,t}$, import-instance fixed effects and year fixed effects.}
\end{table}
Column (1) shows that the estimated coefficient for year $t$ is negative and statistically significant different from zero. In terms of magnitude, a positive one standard deviation import-cost-shock ($24.4\%$ firm level price increase) leads to a reduction in firms' imported quantities of 7.44\% at the product-source country level. However, the impact of an import-cost-shock on imported quantities is short lived. Column (2) shows that an import cost shock in period $t$ does not predict lower imported quantities in period $t+1$. Thus, an import cost shock leads to a transitory one year reduction in firm's imported quantities. This result is in line with the presence of costly to adjust linkages and imperfect substitution across firms' foreign suppliers.

\noindent
\textbf{Benchmark exports results.} Following the analysis on imported quantities, I turn to analyzing the impact of a import cost shocks on firms export performance. To do so, I estimate the following regression
\begin{align} \label{eq:intensive_exports}
    \ln QX_{i,p,d,t} &= \alpha_0 + \gamma_{i,p,d} +\gamma_t+ \beta^{X} \text{Shock}_{i,t} + \Gamma_{i,p,d,t} + \epsilon_{i,p,d,t}
\end{align}
where $\ln QX_{i,p,d,t}$ represents the log quantities exported at the firm $i$, product $p$, destination country $d$ level at year $t+j$ for $j=\{0,1,2\}$; $\gamma_{i,p,d}$ represents firm-product-destination fixed effects, $\gamma_t$ is a time fixed effect, and $\Gamma_{i,p,d,t}$ control variables which include lagged values of the outcome variable and the $\text{Shock}_{i,t}$. Parameter $\beta^{X}$ represents exported quantities semi-elasticity to an import cost shock. 

Table \ref{tab:export_cluster_all} presents estimated values of $\beta^{X}$ for periods $t,t+1$ and $t+2$ in Columns (1) to (3), respectively. A positive import cost shock leads to persistent and significantly lower exported quantities.
\begin{table}[ht]
    \centering
    \caption{Import Cost Shocks \& Export Performance}
    \label{tab:export_cluster_all}
    \small 
\begin{tabular}{lccc}
  & \multicolumn{3}{c}{Log - Quantities Exported - $\ln Q_{i,p,d}$} \\
  & (1) & (2) & (3) \\
  & Period $t$ & Period $t+1$ & Period $t+2$ \\ \hline \hline
$\beta^{X}-\text{Shock}_{i,t}$  & -0.0711** & -0.146*** & -0.118* \\
 & (0.0320) & (0.0410) & (0.0668) \\
 &  &  &  \\
Observations & 168,615 & 107,149 & 69,391 \\ \hline \hline
\multicolumn{4}{c}{\footnotesize Two Way Clustered (Firm - Destination Country) Standard Errors in Parentheses} \\
\multicolumn{4}{c}{\footnotesize *** p$<$0.01, ** p$<$0.05, * p$<$0.1} \\
\end{tabular}
\floatfoot{\footnotesize \textbf{Note:} The regressions are computed controlling for the first and second lagged values of the dependent variable $\ln QX_{i,p,d,t}$, export-instance fixed effects and year fixed effects.}
\end{table}
Import cost shocks cause a negative hump-shaped response of exported quantities. In terms of magnitude, a positive one standard deviation shock ($24.4\%$) leads to a drop in exported quantities of 1.90\% period $t$, 2.44\% in $t+1$ and 2.1\% in $t+2$. These results imply a strong imported-input to export complementarity, as the estimated coefficients in Table \ref{tab:import_cluster} and \ref{tab:export_cluster_all} imply that a reduction in imported quantities of 10\% lead to a reduction in exported quantities between 2.3\% and 4.8\%. This result validates the theoretical prediction that a rise in marginal costs caused by a higher cost of imported inputs leads to a reduction of exported quantities.

The persistence of the impact of an imported cost shock is noteworthy. The fact that the estimated coefficient for period $t+1$ is greater than the one for period $t$  may be driven by lags or temporal considerations in the process of production and exporting. Imported inputs which arrive at customs in year $t$ may be used in the production of goods exported through customs in year $t+1$. Nevertheless, the fact that the initial shock leads to lower exported quantities two years into the future suggest that frictions in the market for suppliers also play a significant role. In other words, the slow and costly to adjust foreign supplier linkages not only cause a drop on impact, but also leads to persistently lower export performance across time. This result provides further evidence on domestic firms' costly to adjust linkages with foreign suppliers.

\noindent
\textbf{Results for high-quality exports.} Lastly, I test the theoretical prediction that exports to consumers with higher love for quality are more sensitive to import cost shocks. However, customs data does not provide direct evidence of either product's quality or consumers love for quality. Thus, to test this theoretical prediction I estimate Equation \ref{eq:intensive_exports} partitioning the sample across two dimensions: (i) product differentiation and (ii) destination countries' income per capita levels.

Argentina's export bundle is primarily composed of commodity-based goods.\footnote{According to UNCTAD commodity dependence report, Argentina is a country which depends heavily on agricultural commodity goods. See, \url{https://unctad.org/webflyer/state-commodity-dependence-2019}} For instance, \cite{bernini2018micro} show that under different good classifications and across time, around 60\% of Argentina's exports are undifferentiated or commodity based goods. A priori, it is expected that products which involve a higher degree of differentiation and customization are of a higher quality and more exposed to imported input cost shocks. To test the model prediction that exports to consumers with higher love for quality are more sensitive to import cost shocks, I partition the sample using two different product classifications: (i) \cite{bernini2018micro} classification between differentiated and undifferentiated goods, (ii) \cite{rauch1999networks} classification between undifferentiated or world priced goods and differentiated goods.\footnote{The good classification introduced by \cite{rauch1999networks} is composed of two classifications, conservative and liberal, into three different type of goods, world-priced goods, reference priced goods and differentiated goods. The results presented in this section follow the liberal good classification although results are robust to using the conservative classification.} The former classification was specifically constructed for the Argentinean case, while the latter classification has been widely used by the literature in international trade.

Table \ref{tab:export_cluster_all_diff} presents estimated values of $\beta^{X}$ for differentiated and undifferentiated goods.
\begin{table}[ht]
    \centering
    \caption{Import Cost Shock \& Export Performance \\ Differentiated Goods}
    \label{tab:export_cluster_all_diff}
    \footnotesize
\begin{tabular}{lcccccc} 
  & \multicolumn{6}{c}{\cite{bernini2018micro} - Log - Quantities Exported - $\ln Q_{i,p,d}$} \\
 & \multicolumn{2}{c}{Period $t$} & \multicolumn{2}{c}{Period $t+1$} & \multicolumn{2}{c}{Period $t+2$} \\
\small & $D$ & $U$ & $D$ & $U$ & $D$ & $U$   \\
 & (1) & (2) & (3) & (4) & (5) & (6) \\ \hline \hline
 $\beta^{X}-\text{Shock}_{i,t}$  & -0.0856** & -0.0518 & -0.168*** & -0.0772 & -0.162** & -0.128 \\
 & (0.0425) & (0.0434) & (0.0581) & (0.158) & (0.0764) & (0.0931) \\
  &  &  &  &  &  &  \\
Observations & 116,743 & 34,324 & 73,359 & 22,340 & 47,013 & 14,866 \\ \hline \hline  
\multicolumn{7}{c}{ Two Way Clustered (Firm - Destination Country) Standard Errors in Parentheses} \\
\multicolumn{7}{c}{ *** p$<$0.01, ** p$<$0.05, * p$<$0.1} \\
\multicolumn{7}{c}{} \\
\end{tabular}
\begin{tabular}{lcccccc} 
  & \multicolumn{6}{c}{\cite{rauch1999networks} -Log - Quantities Exported - $\ln Q_{i,p,d}$} \\
 & \multicolumn{2}{c}{Period $t$} & \multicolumn{2}{c}{Period $t+1$} & \multicolumn{2}{c}{Period $t+2$} \\
\small & $D$ & $U$ & $D$ & $U$ & $D$ & $U$   \\
 & (1) & (2) & (3) & (4) & (5) & (6) \\ \hline \hline 
 $\beta^{X}-\text{Shock}_{i,t}$ & -0.0714** & 0.0136 & -0.159*** & 0.0873 & -0.140* & -0.000858 \\
 & (0.0338) & (0.0687) & (0.0438) & (0.180) & (0.0744) & (0.158) \\
  &  &  &  &  &  &  \\
Observations & 151,035 & 8,953 & 95,795 & 5,733 & 62,024  & 3,686 \\ \hline \hline  
\multicolumn{7}{c}{ Two Way Clustered (Firm - Destination Country) Standard Errors in Parentheses} \\
\multicolumn{7}{c}{ *** p$<$0.01, ** p$<$0.05, * p$<$0.1} \\
\end{tabular}
\floatfoot{\footnotesize \textbf{Note:} The regressions are computed controlling for the first and second lagged values of the dependent variable $\ln QX_{i,p,d,t}$, export-instance fixed effects and year fixed effects.}
\end{table}
The top panel shows the results using the \cite{bernini2018micro} classification, while the bottom panel shows the results using the \cite{rauch1999networks} classification. Across product classifications, a positive import cost shock leads to a persistent drop in exported quantities of differentiated goods but not for undifferentiated goods. On the one hand, the top panel shows that the estimated coefficient for exported quantities of differentiated goods is between 15\% and 37\% greater than those found for the full sample in Table \ref{tab:export_cluster_all}, across time horizons.\footnote{Under the \cite{rauch1999networks} classification, the estimated coefficients are between 0.5\% and 26\% across time periods.} On the other hand, the estimated coefficients for undifferentiated products are not significantly different from zero across any of the time periods. The fact that exported quantities of differentiated goods are more sensitive to import cost shocks than those of undifferentiated goods is in line with the theoretical framework's predictions in Section \ref{sec:theoretical_framework}. This is because there is a higher probability that differentiated goods involve a greater amount of imported inputs and are sold to consumers with higher love for quality than undifferentiated or commodity based goods. 

To provide additional evidence that exports to consumers with higher love for quality are more sensitive to import cost shocks I carry out a sample partition exercise across destination countries. In particular, I partition countries across their income levels based on evidence that ``\textit{High-Income}'' countries demanding higher quality products.\footnote{See for \cite{hallak2011estimating,bastos2018export} for example.} The partition is carried out using the IMF's World Economic Outlook partition between Advanced Economies (Adv.) and Emerging Markets (EM).

The top panel of Table \ref{tab:expo_cluster_all_HI} presents the estimated coefficients for $\beta^X$ of this second sample partition exercise for all exports. For all time periods the estimated semi-elasticity for exported quantities to ``\textit{Advanced-Economies}'' is higher than that estimated for exported quantities to ``\textit{Emerging-Economies}''.
\begin{table}[ht]
    \centering
    \caption{Import Cost Shocks \& Export Performance \\ By Destination Country's Income}
    \label{tab:expo_cluster_all_HI}
    \footnotesize
\begin{tabular}{lcccccc}
  & \multicolumn{6}{c}{Log - Quantities Exported - Full Sample - $\ln Q_{i,p,d}$} \\
 & \multicolumn{2}{c}{Period $t$} & \multicolumn{2}{c}{Period $t+1$} & \multicolumn{2}{c}{Period $t+2$} \\
\small & Adv. & EM & Adv. & EM & Adv. & EM \normalsize \\
 & (1) & (2) & (3) & (4) & (5) & (6) \\ \hline \hline
$\beta^{X}-\text{Shock}_{i,t}$ & -0.173*** & -0.0487 & -0.194** & -0.147*** & -0.171* & -0.119 \\
 & (0.0582) & (0.0326) & (0.0717) & (0.0502) & (0.101) & (0.0787) \\
 &  &  &  &  &  &  \\
Observations & 29,154 & 139,448 & 17,920 & 89,227 & 11,514 & 57,875 \\ \hline \hline
\multicolumn{7}{c}{ Two Way Clustered (Firm - Destination Country) Standard Errors in Parentheses} \\
\multicolumn{7}{c}{ *** p$<$0.01, ** p$<$0.05, * p$<$0.1} \\
\multicolumn{7}{c}{} \\
\end{tabular}
\begin{tabular}{lcccccc}
  & \multicolumn{6}{c}{Log - Quantities Exported - Differentiated Goods - $\ln Q_{i,p,d}$} \\
   & \multicolumn{2}{c}{Period $t$} & \multicolumn{2}{c}{Period $t+1$} & \multicolumn{2}{c}{Period $t+2$} \\
\small & Adv. & EM & Adv. & EM & Adv. & EM \normalsize \\
 & (1) & (2) & (3) & (4) & (5) & (6) \\ \hline \hline
$\beta^{X}-\text{Shock}_{i,t}$ & -0.212*** & -0.0719 & -0.299** & -0.157** & -0.239* & -0.161* \\
 & (0.0619) & (0.0496) & (0.125) & (0.0683) & (0.141) & (0.0862) \\
 &  &  &  &  &  &  \\
Observations & 18,569 & 98,166 & 11,091 & 62,268 & 6,952 & 40,061 \\ \hline \hline
\multicolumn{7}{c}{ Two Way Clustered (Firm - Destination Country) Standard Errors in Parentheses} \\
\multicolumn{7}{c}{ *** p$<$0.01, ** p$<$0.05, * p$<$0.1} \\
\end{tabular}
\floatfoot{\footnotesize \textbf{Note:} The regressions are computed controlling for the first and second lagged values of the dependent variable $\ln QX_{i,p,d,t}$, export-instance fixed effects and year fixed effects. The bottom panel partitions the sample between differentiated and undifferentiated goods using the \cite{bernini2018micro} classification.}
\end{table}
On the one hand, the estimated coefficients for exported quantities to Advanced Economies are negative and statistically significant from zero for all time periods. The estimated semi-elasticities are between $32\%$ and $243\%$ greater than those estimated for the full benchmark sample. These estimated coefficients imply that an import cost shock which decreases imported quantities by 10\%, decreases exported quantities by roughly 6.4\%. On the other hand, the impact of an import cost shock on exported quantities to Emerging Economies is only significantly different from zero for period $t+1$ and with a relatively lower magnitude to that estimated for exports to Advanced Economies.

The bottom panel of Table \ref{tab:expo_cluster_all_HI} presents the estimates of $\beta^{X}$ partitioning the sample between Advanced Economies and Emerging Markets but only for exports of differentiated goods. Within exports of differentiated, those to Advanced Economies exhibit estimated coefficients larger in magnitude compared to those estimated for Emerging Markets across all the time periods considered. This result provides evidence in favor of the model's prediction of exports to consumers with higher love for quality being more sensitive to import cost shocks. 

\section{Conclusion} \label{sec:conclusion}

This paper exploits information on firms' foreign supplier network to analyze how increases in the cost of imported inputs affects their export performance. The constructed theoretical framework predicts that the impact of an increase in the cost of imported inputs is increasing in the adjustment costs of firms' linkages and in the love for quality of destination countries.

Motivated by these predictions and based on the theoretical framework' search frictions in the market for foreign input suppliers, I construct firm-specific shocks using a shift share methodology. These shocks exploit firms' lagged exposure to foreign suppliers and finely defined import price shifts. I find that a positive import cost shock leads to lower imported quantities, providing evidence that firms face costly to adjust linkages. Furthermore, an import cost shocks leads to significant and persistent drop in exported quantities up to two years after the initial shock. Exports of differentiated products and exports to high-income countries exhibit an estimated semi-elasticity twice as big as the benchmark results in line with the model's predictions. 

This paper shed light on the inherent frictions in the market for imported inputs. Results suggest the presence of significant adjustment costs in domestic firms' foreign supplier linkages, and provide evidence of strong complementarities between imported inputs and exports. These results are of particular importance for Emerging Market economies which seek to diversify export bundles whilst lacking developed networks of domestic suppliers.

\newpage
\bibliography{biblio.bib}

\newpage
\appendix

\section{Appendix} \label{sec:Appendix}

\subsection{Model Details} \label{subsec:appendix_model_details}

This section of the appendix presents additional details on the model presented in Section \ref{sec:theoretical_framework}. I start by describing firm $i$'s production line $d$ profit maximization problem

\subsubsection{Firms' Profit Maximization \& FOCs} \label{subsubsec:appendix_profit_maximization_FOC}

\noindent
\textbf{Profit Maximization.} Firm $i$'s production line $d$'s profit maximization problem can be stated as
\begin{equation*}
\begin{aligned}
\max_{\lambda_{i,d},x_{i,d},p_{i,d}} \pi_{i,d} \left(z_i, \xi_i, c_{j} \right) = \quad & \left(p_{i,d} - C_{i} \right) x_{i,d} - \frac{f}{\xi_i}\lambda_{i,d} - F^{d} \mathbbm{1}\left[x_{i,d}>0\right] \\ 
\textrm{s.t.} \quad & x_{i,d} = \frac{\left(p_{i,d}\right)^{-\rho} \left(\lambda_{i,d}\right)^{\left(\rho-1\right) \zeta_d} y_d }{P^{-\rho}_d} \\
  &  C_{i} \left(z_i, c_{j} \right) = \frac{1}{z_{i}} \frac{1}{\alpha^{\alpha}(1-\alpha)^{1-\alpha}} w^{\alpha} \left(c_j\right)^{1-\alpha}  \\
\end{aligned}
\end{equation*}
Both constraints can be introduced inside the objective function. Consequently, the objective function can be re-written as
\begin{equation}
\begin{aligned}
\max_{\lambda_{i,d},x_{i,d}} \pi_{i,d} \left(z_i, \xi_i, \lambda_{i,d} \right) = \quad & x^{1-\frac{1}{\rho}}_{i,j} y^{\frac{1}{\rho}}_j P_j \left(\lambda_j\right)^{\frac{\left(\rho-1\right)\zeta_j}{\rho}} -  C_{i} x_{i,d}  - \frac{f}{\xi_i}\lambda - F^{d} \mathbbm{1}\left[x_{i,d}>0\right] \label{eq:profit_maximization_subst_one} \\
\end{aligned}
\end{equation}
where I used an expression for $p_{i,j}$
\begin{equation*}
    p_{i,d} = x^{\frac{-1}{\rho}}_{i,d} y^{\frac{1}{\rho}}_d P_d \left(\lambda_d\right)^{\frac{\left(\rho-1\right)\zeta_d}{\rho}}
\end{equation*}
This problem leads to the following first order conditions on $x_{i,d}$ and $\lambda_{i,d}$
\begin{align}
    \frac{\partial \pi }{\partial x_{i,d}}: & \quad \frac{\rho-1}{\rho} x^{\frac{-1}{\rho}}_{i,d} y^{\frac{1}{\rho}}_d P_d \left(\lambda_{i.d}\right)^{\frac{\left(\rho-1\right)\zeta_d}{\rho}} = C_{i} \left(z_i, c_j\right)  \label{eq:FOC_quantities_multiple_one} \\
    \frac{\partial \pi }{\partial \lambda_{i,d}}: & \quad x^{1-\frac{1}{\rho}}_{i,d} y^{\frac{1}{\rho}}_d P_d \left(\lambda_d\right)^{\frac{\left(\rho-1\right)\zeta_d}{\rho}-1} \left(\frac{\rho-1}{\rho} \right) \zeta_d = \frac{f}{\xi_i} \label{eq:FOC_quality_one}
\end{align}
Next, I solve for a closed form solution of the endogenous variables $\{x_{i,d},\lambda_{i,d}\}$. Only the endogenous variables and $c_j$ are of interest, so I agglomerate all other exogenous variables into constants. I re-write the system of equations presented by Equations \ref{eq:FOC_quantities_multiple_one} and \ref{eq:FOC_quality_one} as
\begin{align}
    \frac{\partial \pi }{\partial x_{i,d}}: & \quad x^{-\frac{1}{\rho}}_{i,d} \lambda^{\left(\frac{\rho-1}{\rho}\right)\zeta_d}_{i,d} = K_1 C_i \label{eq:reformulated_x}\\
    \frac{\partial \pi }{\partial \lambda_{i,d}}: & \quad x^{1-\frac{1}{\rho}}_{i,d} \lambda^{\left(\frac{\rho-1}{\rho}\right)\zeta_d-1}_{i,d} = K_2 \label{eq:reformulated_lambda}
\end{align}
where
\begin{align*}
    K_1 &= \left[\left(\frac{\rho-1}{\rho} y^{\frac{1}{\rho}}_d P_d\right) \right]^{-1} \\
    K_2 &= \left[\frac{\xi}{f} y^{\frac{1}{\rho}}_d P_d \left(\frac{\rho-1}{\rho}\right) \zeta_d  \right]
\end{align*}
Dividing Equation \ref{eq:reformulated_lambda} by Equation \ref{eq:reformulated_x}, we obtain
\begin{align}
    x_{i,d}  = \frac{K_2}{K_1 C_i} \lambda_{i,d}
\end{align}
Replacing this expression for $x_{i,d}$ on Equation \ref{eq:reformulated_x} we can solve for optimal quality $\lambda_{i,d}$
\begin{align}
    \left(\frac{K_2}{K_1 C_i} \lambda_{i,d}\right)^{-\frac{1}{\rho}} \lambda^{\left(\frac{\rho-1}{rho}\right)\zeta_d}_{i,d} = K_1 C_i
\end{align}
\begin{align}
    \lambda^{*}_{i,d} &= \left[K^{\frac{\rho-1}{\rho}}_1 K_2 C^{\frac{\rho-1}{\rho}}_i \right]^{\frac{\rho}{\left(\rho-1\right)\zeta_d-1}} \\
    &= \left[K^{\frac{\rho-1}{\rho}}_1 K_2 \right]^{\frac{\rho}{\left(\rho-1\right)\zeta_d-1}} C^{\frac{\rho-1}{\left(\rho-1\right)\zeta_d-1}}_i
\end{align}
Grouping together the terms that do not depend on $C_i$, optimal quality can be expressed as
\begin{align}
    \lambda^{*}_{i,d} &= \Lambda_1 \times C^{\frac{\rho-1}{\left(\rho-1\right)\zeta_d-1}}_i
\end{align}
where $\Lambda_1 = K^{\frac{\rho-1}{\left(\rho-1\right)\zeta_d-1}}_1 \times K^{\frac{\rho}{\left(\rho-1\right)\zeta_d-1}}_2$.

\noindent
Next, I solve for the optimal quantities sold as
\begin{align}
    x^{*}_{i,d} &= \frac{K_2}{K_1 C_i} \lambda^{*}_{i,d} \\
                &= \frac{K_2}{K_1} C^{-1}_i K^{\frac{\rho-1}{\left(\rho-1\right)\zeta_d-1}}_1 K^{\frac{\rho}{\left(\rho-1\right)\zeta_d-1}}_2 C^{\frac{\rho-1}{\left(\rho-1\right)\zeta_d-1}}_i \\
                &= K^{\frac{\left(1-\zeta_d\right)\left(\rho-1\right)-1}{\left(\rho-1\right)\zeta_d-1}}_1 K^{\frac{\rho+\zeta_d\left(\rho-1\right)-1}{\left(\rho-1\right)-1}}_2 C^{\frac{\rho-1}{\left(\rho-1\right)\zeta_d-1}-1}_i 
\end{align}
Grouping together the terms that do not depend on $C_i$, optimal quantities sold can be expressed as
\begin{align}
    x^{*}_{i,d} &= \Lambda_2 \times  C^{\frac{\left(1-\zeta_d\right)\left(\rho-1\right)-1}{\left(\rho-1\right)\zeta_d-1}}_i
\end{align}
where $\Lambda_2 =  K^{\frac{\left(1-\zeta_d\right)\left(\rho-1\right)-1}{\left(\rho-1\right)\zeta_d-1}}_1 K^{\frac{\rho+\zeta_d\left(\rho-1\right)-1}{\left(\rho-1\right)-1}}_2$.

\subsubsection{Firms' Profit Maximization \& SOCs} \label{subsubsec:appendix_profit_maximization_SOC}

Starting from the profit maximization problem
\begin{equation*}
\begin{aligned}
\max_{\lambda_{i,d},x_{i,d}} \pi_{i,d} \left(z_i, \xi_i, \lambda_{i,d} \right) = \quad & x^{1-\frac{1}{\rho}}_{i,j} y^{\frac{1}{\rho}}_j P_j \left(\lambda_j\right)^{\frac{\left(\rho-1\right)\zeta_j}{\rho}} -  C_{i} x_{i,d}  - \frac{f}{\xi_i}\lambda - F^{d} \mathbbm{1}\left[x_{i,d}>0\right] \\
\end{aligned}
\end{equation*}
and the first order conditions
\begin{align*}
    \frac{\partial \pi }{\partial x_{i,d}}: & \quad \frac{\rho-1}{\rho} x^{\frac{-1}{\rho}}_{i,d} y^{\frac{1}{\rho}}_d P_d \left(\lambda_{i.d}\right)^{\frac{\left(\rho-1\right)\zeta_d}{\rho}} = C_{i} \left(z_i, c_j\right)   \\
    \frac{\partial \pi }{\partial \lambda_{i,d}}: & \quad x^{1-\frac{1}{\rho}}_{i,d} y^{\frac{1}{\rho}}_d P_d \left(\lambda_d\right)^{\frac{\left(\rho-1\right)\zeta_d}{\rho}-1} \left(\frac{\rho-1}{\rho} \right) \zeta_d = \frac{f}{\xi_i}
\end{align*}
for the second order conditions to be met I need to prove that
\begin{align*}
    \frac{\partial^2 \pi}{\partial x^2 _{i,d}} &<0 \\
    \frac{\partial^2 \pi}{\partial \lambda^2 _{i,d}} &<0 \\
    \frac{\partial^2 \pi}{\partial x^2 _{i,d}} \frac{\partial^2 \pi}{\partial \lambda^2 _{i,d}} & - \left(\frac{\partial^2 \pi}{\partial x_{i,d} \lambda_{i,d}}\right)^2 > 0
\end{align*}

\noindent
First, proving that $\frac{\partial^2 \pi}{\partial x^2 _{i,d}} <0$ is straightforward as
\begin{align*}
    \frac{\partial^2 \pi}{\partial x^2 _{i,d}} = -\frac{1}{\rho} x^{-\frac{1}{\rho}-1}_{i,d} \times \lambda^{\frac{\left(\rho-1\right)\zeta_d}{\rho}}_{i,d} \left[\frac{\rho-1}{\rho} y^{\frac{1}{\rho}}_d P_d \right] < 0
\end{align*}
as $\rho>1$ and $-1/\rho < 0$. 

\noindent
Second, I need to determine under which conditions $\frac{\partial^2 \pi}{\partial \lambda^2 _{i,d}} <0$. This second derivative is given by
\begin{align*}
    \frac{\partial^2 \pi}{\partial \lambda^2 _{i,d}} =  x^{1-\frac{1}{\rho}}_{i,d} \times \lambda^{\frac{\left(\rho-1\right)\zeta_d-2}{\rho}}_{i,d} 
    \zeta_d \left[\frac{\left(\rho-1\right)}{\rho} \zeta_d -1 \right] 
    \left[\frac{\rho-1}{\rho} y^{\frac{1}{\rho}}_d P_d \right]
\end{align*}
Given that $x_{i,d},\lambda_{i,d},\zeta_d>0$ and $\rho>1$, for $\frac{\partial^2 \pi}{\partial \lambda^2 _{i,d}} <0$, a necessary condition is that
\begin{align} \label{eq:SOC_necessary_condition}
    \frac{\left(\rho-1\right)}{\rho} \zeta_d -1  < 0 
\end{align}
which implies
\begin{align} \label{eq:SOC_necessary_condition_constraint}
    \zeta_d < \frac{\rho}{\rho-1}
\end{align}
Note that if $\frac{\left(\rho-1\right)}{\rho} \zeta_d -1  < 0 $ is satisfied, then condition $\left(\rho-1\right) \zeta_d -1  < 0 $ is also satisfied as $\rho>1$.

\noindent
The partial derivatives $\frac{\partial^2 \pi}{\partial x_{i,d} \lambda_{i,d}}$ are equal to
\begin{align*}
    \frac{\partial^2 \pi}{\partial x_{i,d} \lambda_{i,d}} = \left[ \frac{\rho-1}{\rho} y^{\frac{1}{\rho}}_d P_d \right] \left(\frac{\rho-1}{\rho}\right) \zeta_d x^{-\frac{1}{\rho}}_{i,d} \lambda^{\left(\frac{\rho-1}{\rho}\right)\zeta_d -1}
\end{align*}
which is strictly positive as $\rho>1$.

\noindent
Next, I turn to test under which conditions the last of the second order conditions of the profit maximization problem are met. This is
\begin{align*}
    \frac{\partial^2 \pi}{\partial x^2 _{i,d}} \frac{\partial^2 \pi}{\partial \lambda^2 _{i,d}}  - \left(\frac{\partial^2 \pi}{\partial x_{i,d} \lambda_{i,d}}\right)^2 > 0
\end{align*}
To begin with, I compute the square of the cross partial derivatives
\begin{align*}
    \left(\frac{\partial^2 \pi}{\partial x_{i,d} \lambda_{i,d}}\right)^2 = \left[ \frac{\rho-1}{\rho} y^{\frac{1}{\rho}}_d P_d \right]^2 \left(\frac{\rho-1}{\rho}\right)^2 \zeta^2_d x^{-\frac{2}{\rho}}_{i,d} \lambda^{\left(2\frac{\rho-1}{\rho}\right)\zeta_d -2}
\end{align*}
Next, I compute the product of the second order derivatives of the profit function with respect to $x_{i,d}$ and $\lambda_{i,d}$
\begin{align*}
    \frac{\partial^2 \pi}{\partial x^2 _{i,d}} \frac{\partial^2 \pi}{\partial \lambda^2 _{i,d}} &= \left[ \frac{\rho-1}{\rho} y^{\frac{1}{\rho}}_d P_d \right]^2 \times \left(-\frac{1}{\rho}\right) \zeta_d \times \left[\left(\frac{\rho-1}{\rho}\right)\zeta_d-1 \right] \times x^{-\frac{2}{\rho}}_{i,d} \times \lambda^{2\left(\frac{\rho-1}{\rho}\right)\zeta_d - 2}_{i,d}
\end{align*}
The expressions for $\left(\frac{\partial^2 \pi}{\partial x_{i,d} \lambda_{i,d}}\right)^2$ and $\frac{\partial^2 \pi}{\partial x^2 _{i,d}} \frac{\partial^2 \pi}{\partial \lambda^2 _{i,d}}$ share the terms
\begin{align*}
    \left[ \frac{\rho-1}{\rho} y^{\frac{1}{\rho}}_d P_d \right]^2 \times x^{-\frac{2}{\rho}}_{i,d} \times \lambda^{2\left(\frac{\rho-1}{\rho}\right)\zeta_d - 2}_{i,d} \zeta_d
\end{align*}
which can be factored out. Thus, for condition $\frac{\partial^2 \pi}{\partial x^2 _{i,d}} \frac{\partial^2 \pi}{\partial \lambda^2 _{i,d}}  - \left(\frac{\partial^2 \pi}{\partial x_{i,d} \lambda_{i,d}}\right)^2 > 0$ to be satisfied, it must be the case that
\begin{align*}
    \Bigg\{\left(-\frac{1}{\rho}\right)\left[\left(\frac{\rho-1}{\rho}\right)\zeta_d-1 \right]-\left(\frac{\rho-1}{\rho}\right)^2 \zeta_d  \Bigg\} > 0
\end{align*}
This condition can be re-written as
\begin{align*}
    \Bigg\{ \frac{-\left(\rho-1\right)\zeta_d + \rho}{\rho^2} - \frac{\left(\rho-1\right)^2 \zeta_d}{\rho^2} \Bigg\} >0
\end{align*}
taking the denominator $\rho^2$ out of the brackets as $\rho>1$, the condition now becomes
\begin{align*}
    -\left(\rho-1\right)\zeta_d + \rho - \left(\rho-1\right)^2 \zeta_d > 0
\end{align*}
Grouping together the terms $\left(\rho-1\right)\zeta_d$, the condition becomes
\begin{align*}
    & \left(\rho-1\right)\zeta_d \left[-1 - \left(\rho-1\right) \right] + \rho>0 \\
    & \left(\rho-1\right)\zeta_d \times \left[-\rho \right] + \rho>0
\end{align*}
Which finally can be expressed as
\begin{align*}
    \rho - \left(\rho-1\right)\zeta_d \times \rho > 0 \\
    \rho \left(1 - \left(\rho-1\right)\zeta_d\right) > 0
\end{align*}
Then, as long as
\begin{align*}
    \left(\rho-1\right)\zeta_d < 1
\end{align*}
The second order conditions are satisfied. However, note that this condition is less binding that the condition necessary for the second order derivative $\frac{\partial^2 \pi}{\partial \lambda^2 _{i,d}}$ to be satisfied. 

\newpage
\subsubsection{Model Features} \label{subsubsec:appendix_model_features}

In this subsection of the Appendix I carry out numerical simulation results to analyze in detail the economics of the model presented in Section \ref{sec:theoretical_framework}. The goal of these numerical exercises is not to quantitatively match all features of the data but to display the model's main mechanisms at work. 

For simplicity, I assume that there are only two destination export markets, a relatively richer destination country, denoted country "$R$", and a relatively poorer destination country, denoted country "$P$". Additionally, I introduce a domestic market which entails no fixed costs of exporting which I label country "$D$".\footnote{The introduction of the domestic economy is not innocuous. A firm might be active selling in the domestic market alone because the low efficiency of their current supplier prevents them from exporting. Matching with a new supplier may both allow them to enter export markets and increase their profits in the domestic market. Consequently, the incentives to search are going to be higher than if only export markets are considered.} The time frequency is annual to match the data used in previous sections. To begin with, I describe firm's export performance for different levels of idiosyncratic variables $\{z_i,\xi_i\}$ and for different values of foreign supplier efficiency $c_j$.

\begin{figure}[ht]
    \caption{Firms' export scope according to $c_j$}
    \label{fig:heatmap}
     \centering
     \begin{subfigure}[b]{0.495\textwidth}
         \centering
         \includegraphics[width=\textwidth]{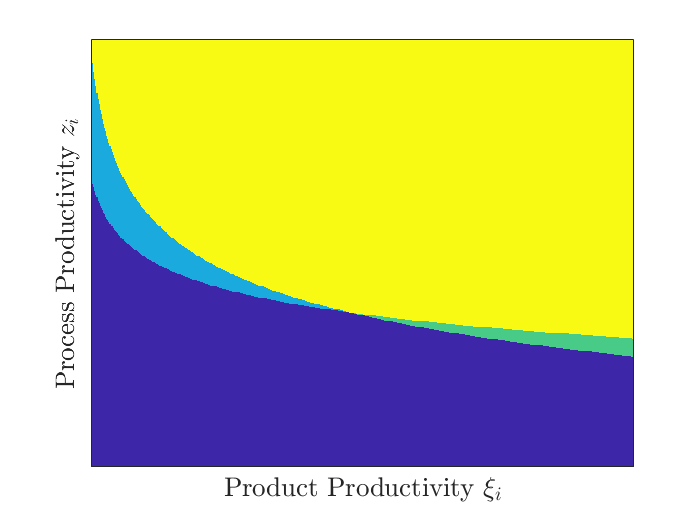}
         \caption{Low $c_j$}
         \label{fig:HeatMap_HighFE}
     \end{subfigure}
     \hfill
     \begin{subfigure}[b]{0.495\textwidth}
         \centering
         \includegraphics[width=\textwidth]{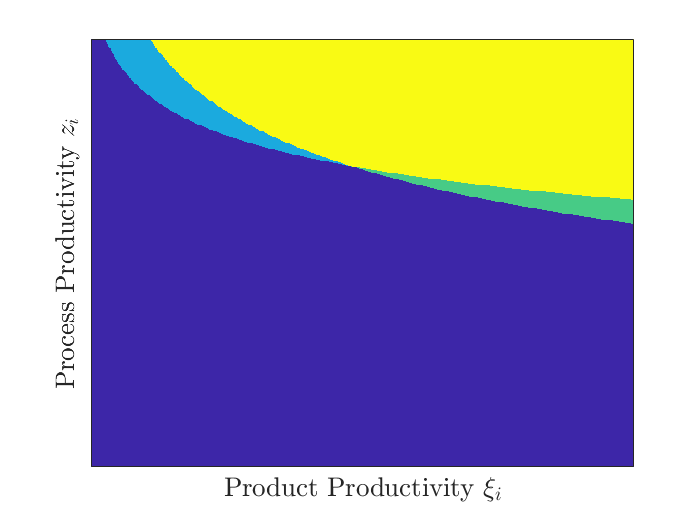}
         \caption{High $c_j$}
         \label{fig:HeatMap_LowFE}
     \end{subfigure}
    \floatfoot{\footnotesize \textbf{Note:} The data on the figures are generated by simulating 250,000 firms with different combinations of $z_i$ and $\xi_i$. The dark blue are denotes firms which do not export and only serve the domestic market. The light blue area denotes firms which export only to ``Country $P$''. The green area denotes firms which export only to  ``Country $R$''. The yellow area denotes firms which export to both countries.}
\end{figure}

I start by showing how firms' export scope depend on both firms' $\{z_i, \xi_i \}$ and foreign supplier's efficiency. Figure presents the result. The left panel or Figure \ref{fig:HeatMap_HighFE} shows the firms' export scope when matched with a high-efficiency foreign supplier, i.e., low $c_j$. The model shows that a large share of firms do not export at all (dark blue area), a subset of firms export only to one destination (``Country $P$'' and ``Country $R$''), and another subset of firms exports to both destination countries simultaneously. 
Firms with both low $\{z_i, \xi_i \}$ do not export at all and just serve the domestic market (dark blue area). Firms with a moderately high $z_i$ but low $\xi_i$ are only able to export to ``Country $P$'' as they are not able to produce a significantly high-quality product to profitably serve ``Country $R$'' (light blue area). Firms with a low $z_i$ but moderately high $\xi_i$ are only able to export to ``Country $R$'' as they can produce high-quality cheaply, but can not produce at a sufficiently low marginal cost to profitably serve ``Country $P$'' (green area). Firms with high $z_i$ and $\xi_i$ are able to export simultaneously to both destination countries (yellow area). The right panel or Figure \ref{fig:HeatMap_LowFE} shows firms' export scope for a higher $c_j$, i.e., when firms are matched with a foreign supplier with lower efficiency. This increase in the marginal cost of production, or $C_i$, leads to a reduction of firms' export scope as the figure suggest that a measure of firms exit one or even both export markets. Note that firms with lower $z_i$ or $\xi_i$ are more susceptible to exit export markets, in line with the empirical results presented in Section \ref{sec:results}.

\begin{figure}[ht]
    \centering
    \caption{Firms' search decisions according to $\{z_i, \xi_i\}$ \\ Static Exercise}
    \label{fig:HeatMap_SearchEfficiency}
    \includegraphics[scale=0.5]{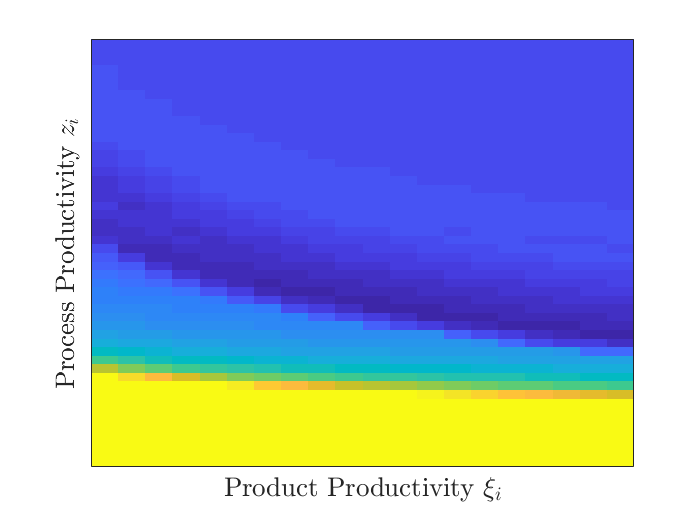}
    \floatfoot{\footnotesize \textbf{Note:} The data on the figures are generated by simulating 50,000 firms with different combinations of $\{z_i,\xi_i,c_j\}$. Each cell of the heatmaps represents a value of $c_j$, with higher values graphed in yellow colors, intermediate values fading from orange to clear blue colors, and low values graphed in dark blue colors.}
\end{figure}

Second, I show how firms' decision to search for foreign suppliers depends on $z_i$ and $\xi_i$ under the current parametrization. Figure \ref{fig:HeatMap_SearchEfficiency} presents a heatmap firm $i$'s search threshold of $c_j$, i.e., the foreign supplier efficiency at which firm $i$ is indifferent between searching or not.
Firms with low $z_i$ and low $\xi_i$ do not search for foreign suppliers, even with $c_j = \bar{c}$ (graphed in bright yellow). This is also the case for firms with low $z_i$ but high $\xi_i$ (at least under our parametrization). This is driven by $z_i$ and $c_j$ affecting export profits through the overall production marginal cost $C_i$. Thus, for significantly low $z_i$, large values of $\xi_i$ are not enough to make the expected value of searching for a higher $c_j$ higher than the fixed cost associated with searching. As $z_i$ increases firms start having lower values of $c_j$ than $\bar{c}$ graphed in colors ranging from orange, green to light blue. The lowest values of threshold $c_j$ are met for intermediate values of $z_i$ and $\xi_i$. These are the firms which have the highest to gain from searching for high efficiency foreign suppliers. This is relative lower $c_j$ do not affect significantly $C_i$ for firms with significantly $z_i$, thus, not altering their total profits. However, for firms with intermediate levels of $z_i$, searching and matching with a foreign supplier with low $c_j$ will be the difference between exporting to two or one of the countries, or even exporting at all. Lastly, firms with both high $z_i$ and $\xi_i$ also have low threshold values for $c_j$. 

Third, I show how search frictions matter for the dynamics of the matched foreign supplier efficiency. I simulate the path of 1,000 firms' matched foreign supplier efficiency for $T = 1000$ periods, starting from a common $c_j = \bar{c}$.
\begin{figure}
    \caption{Firms' search decisions according to $\{z_i, \xi_i\}$ \\ Dynamic Exercise}
    \label{fig:heatmap_search}
     \centering
     \begin{subfigure}[b]{0.495\textwidth}
         \centering
         \includegraphics[width=\textwidth]{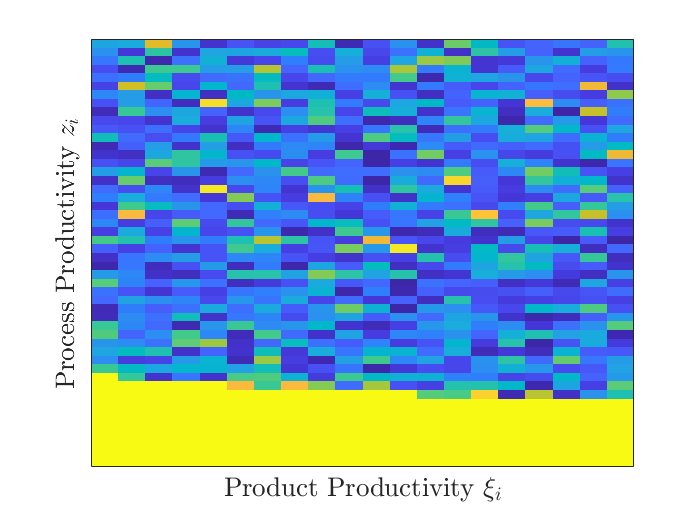}
         \caption{Distribution $c_j$ - $T=5$}
         \label{fig:HeatMap_SearchEfficiency_T5}
     \end{subfigure}
     \hfill
     \begin{subfigure}[b]{0.495\textwidth}
         \centering
         \includegraphics[width=\textwidth]{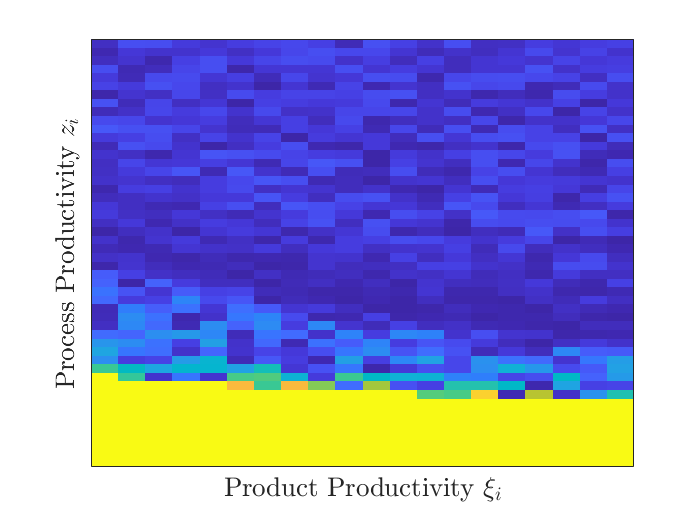}
         \caption{Distribution of $c_j$ - $T=1000$}
         \label{fig:HeatMap_SearchEfficiency_T1000}
     \end{subfigure}
    \floatfoot{\footnotesize \textbf{Note:} The data on the figures are generated by simulating 1,000 firms with different combinations of $z_i$ and $\xi_i$. Each cell of the heatmaps represents a value of $c_j$, with higher values graphed in yellow colors, intermediate values fading from orange to clear blue colors, and low values graphed in dark blue colors.}
\end{figure}
Figure \ref{fig:HeatMap_SearchEfficiency_T5} shows the distribution of firms' matched $c_j$ in period $T=5$. As in Figure \ref{fig:HeatMap_SearchEfficiency}, firms with low $z_i$ do not search for new foreign suppliers and are stuck with $c_j = \bar{c}$. The rest of the firms do search, with no visible pattern emerging after only five periods. Figure \ref{fig:HeatMap_SearchEfficiency_T1000} shows the distribution of firms' matched $c_j$ in period $T=1,000$ showing a clearer pattern. Similar to Figure \ref{fig:HeatMap_SearchEfficiency}, for those firms which search, those with relatively low $z_i$ and $\xi_i$ exhibit higher values of $c_j$ and firms with relatively higher $z_i$ and $\xi_i$. While some firms with low $\{z_i, \xi_i\}$ are matched with high efficiency foreign suppliers, on average this firms stop searching for foreign suppliers faster than firms with higher $\{z_i, \xi_i\}$.

\newpage
\subsection{Cleaning procedure of supplier identification} \label{subsec:Appendix_cleaning}

\noindent In this appendix, I explain the algorithm to clean suppliers' name variable. This is the same algorithm used in \cite{berninietal2021}. Most of the problems detected were typing errors or different ways to write the same firm. For example, "Volkswagen" could be find as (a) "Volkswagen", (b) "Volkswagen international", (c) "VW", (d) "Volkswgen", etc. If I ignore this problem, I could interpret them as 4 different suppliers, when they actually are the same company.

\noindent In order to fix this problem, I apply the following procedure:

\begin{itemize}
    \item $1^{st}$ step: I delete special characters (e.g. "/" and "-"), double spaces, country names (e.g. United States) and their acronym (e.g. USA), and companies' suffixes (LLC, LTD, SA).
    \item $2^{nd}$ step: If a suppliers' name includes the name of a famous multinational firm (or its acronym), I replace the supplier name with them. For example, if a supplier includes Volkswagen or VW, I define them as Volkswagen.
    \item $3^{rd}$ step: I compare the similarity of importers and suppliers names using \cite{raffo2009play} bigram technique (I explain in detail below). If this technique report high similarity, I replace suppliers' name by the importer's name.
    \item $4^{th}$ step: I applied the same procedure of step 3 to compare all suppliers within each firm and determine which are the same company.
\end{itemize}

\noindent The $1^{st}$ step of the procedure allows as to fix cases as previous (b) example by only deleting common words that are not the core name of the company. $2^{nd}$ step is design to solve (c) problems, where a commonly known firm could be rightly write by different ways. 

\noindent However, typing errors, as the example (d), are more difficult to detect and fix. I use \cite{raffo2009play} bigram technique, which compare two string variables and split them by a sequence of two characters called "bigram set" (e.g. "Ford" bigram set is [Fo, or, rd]). The algorithm compute a similscore as the ratio between the union of both bigram sets and all components of both sets. In addition, I use two different weights to account to common bigrams: Simple and log weights. 

\noindent Hence, in $3^{rd}$ and $4^{th}$ step I obtain two similscores in each step, then I define that two strings has "high similarity" if both scores are above 0.65, or one is above 0.8 and the other above 0.35.

\newpage
\subsection{Additional Data Sources} \label{subsec:appendix_additional_data}

I complement my international trade dataset with additional datasets on aggregate macroeconomic variables and international trade for a panel of countries. In Section \ref{sec:results} I use the IMF's World Economic Outlook definition of Advanced Economies and Emerging Markets to proxy export destination countries' proxy of level of income and market sophistication. Additionally, I source CPI and nominal exchange rate from the IMF's IFS dataset to construct measures of bilateral real exchange rates. In Section \ref{sec:results}, I partition the sample across different types of goods. To carry out this partition I use two good classifications: (i) \cite{rauch1999networks} good classification according to whether goods have reference prices in international markets, (ii) \cite{bernini2018micro} good classification according to its degree of differentiation. These good classifications have been used previously in the literature in international economics.\footnote{To access the good classification presented in \cite{rauch1999networks} visit \url{https://econweb.ucsd.edu/~jrauch/rauch_classification.html}, and to access the data set presented by \cite{bernini2018micro} visit \url{https://drive.google.com/file/d/0B19niEgxbWCrUTM3bHNodU9Lenc/view?resourcekey=0-PcCPLyEZGgzRA_GzRjJnvg}.}

\newpage
\subsection{Additional Evidence on the Role of Imported Inputs} \label{subsec:appendix_additional_evidence}

In this section of the Appendix I present additional evidence on the importance of imported inputs at the aggregate and firm level. 

\begin{figure}[ht]
    \centering
    \caption{Relevance of Imported Inputs for Production }
    \label{fig:impo_introduction}
    \begin{subfigure}[b]{0.45\textwidth}
    \centering    
    \includegraphics[scale=0.5]{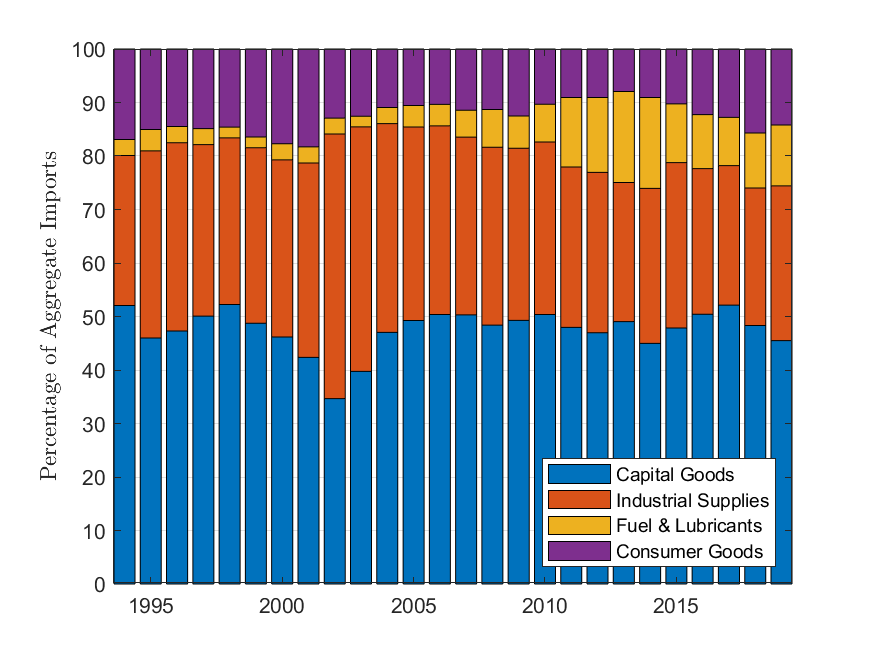}
    \caption{Import decomposition - Main end use}
    \label{fig:impo_bec}
     \end{subfigure}
     \hfill
    \begin{subfigure}[b]{0.45\textwidth}
    \centering
    \includegraphics[scale=0.5]{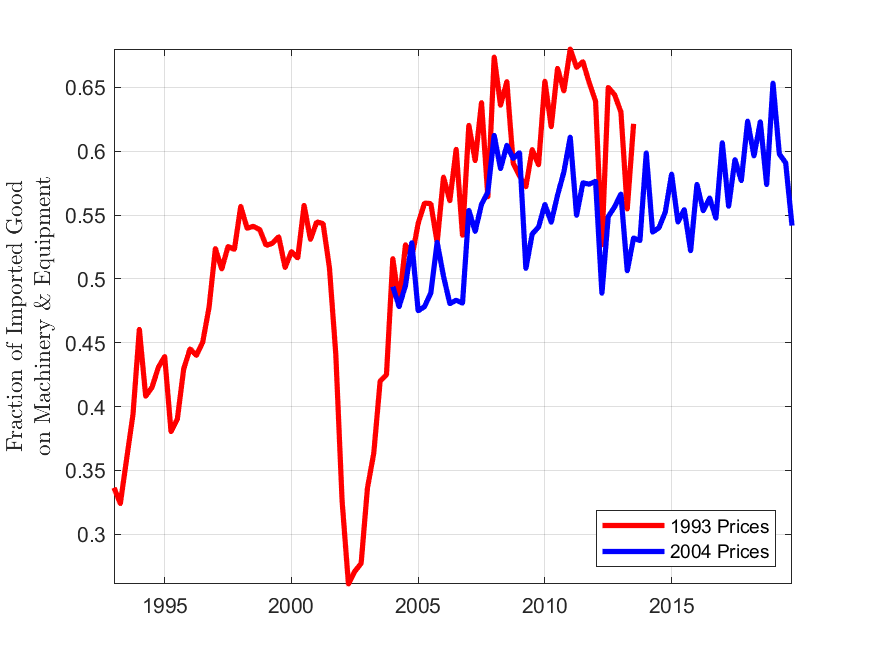}
    \caption{Share of Imports in Investment}
    \label{fig:impo_investment}
     \end{subfigure}
     \floatfoot{\footnotesize \textbf{Note:} The left panel presents the decomposition of total imported value according to the main end use of goods, using the latest revision of the Broad Economic Category (BEC) classification. The right panel presents data on the fraction of imported goods on total investment in capital, sourced from Argentina's national accounts.}
\end{figure}

The reliance of Argentinean firms on foreign intermediate and capital goods can be observed at the aggregate level in Figure \ref{fig:impo_introduction}. On the left panel, Figure \ref{fig:impo_bec} shows that the import of intermediate and capital goods explain between 80\% and 90\% of Argentina's total imports for the period 1994-2019. On the right panel, Figure \ref{fig:impo_investment} shows that imported capital goods explain between 55\% and 65\% of the total investment in machinery and/or equipment for the same period.

\newpage
\subsection{Role of Imports on Exporting} \label{subsec:appendix_import_for_exporting}

In this appendix I present additional evidence on the importance of imports on export performance. First, the regression analysis presented in Section \ref{sec:results} only focuses on exporting firms which imports. Thus, to assert that the results are important for aggregate exports as a whole it is necessary to show the relevance of importing in exporting.

\begin{figure}[ht]
    \caption{Importance of Imports for Exporting}
    \label{fig:import_for_exporting}
     \centering
     \begin{subfigure}[b]{0.495\textwidth}
         \centering
         \includegraphics[width=\textwidth]{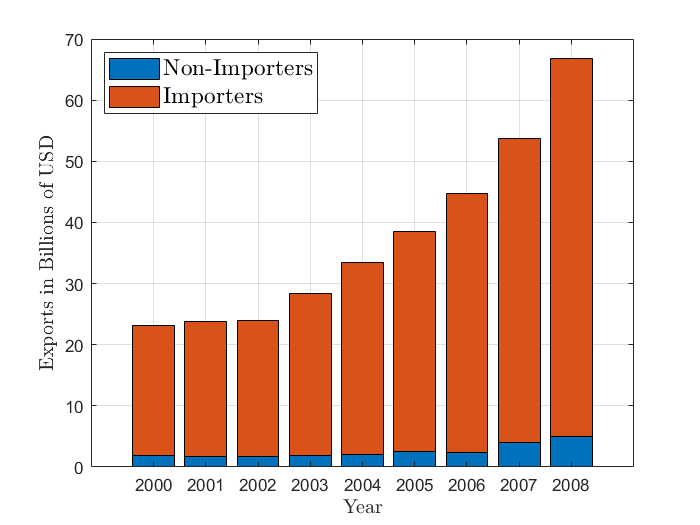}
         \caption{Share of Exports by Importers}
         \label{fig:Export_Decomposition_Importers}
     \end{subfigure}
     \hfill
     \begin{subfigure}[b]{0.495\textwidth}
         \centering
         \includegraphics[width=\textwidth]{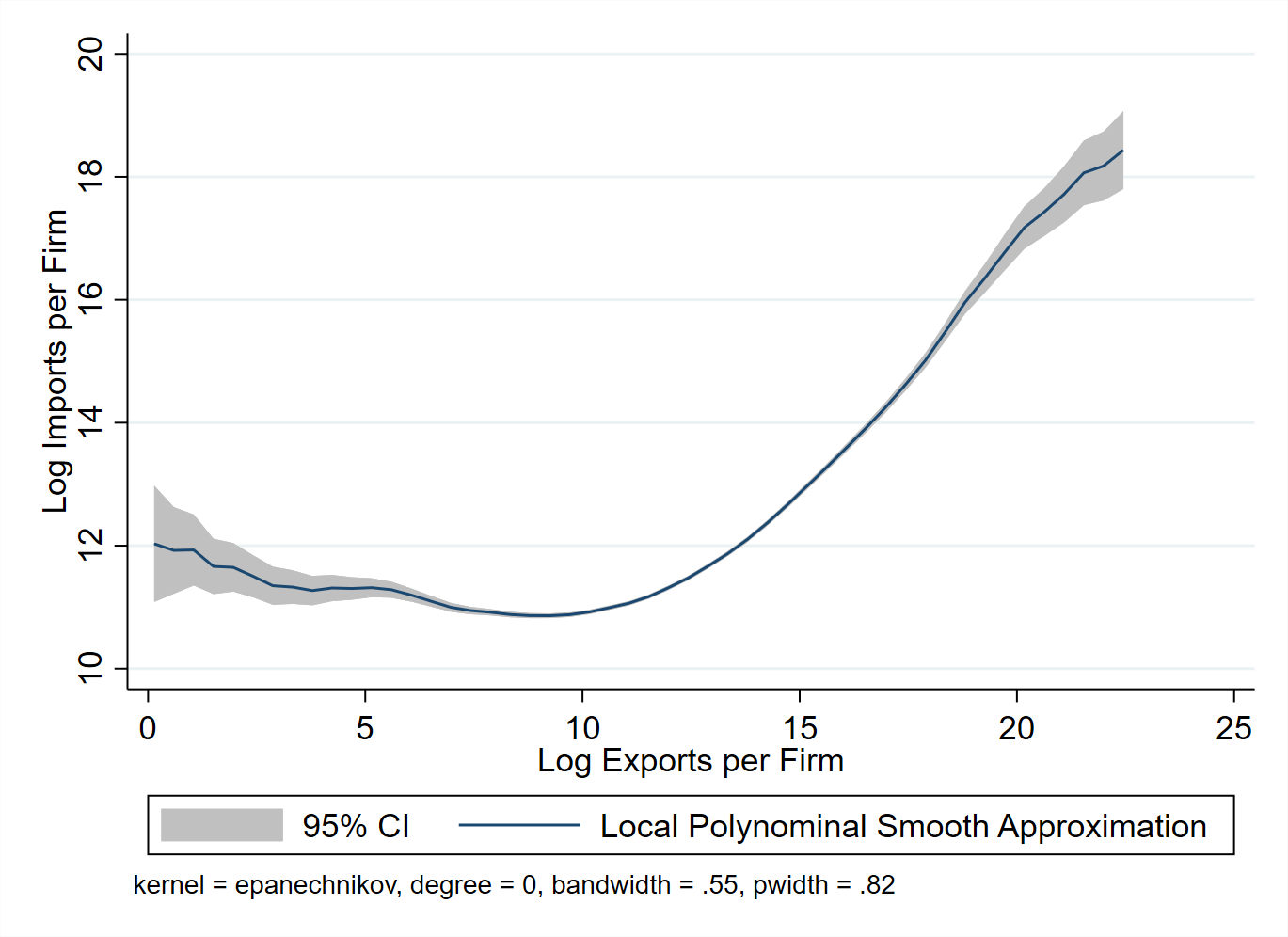}
         \caption{Exports \& Imports per Firm}
         \label{fig:exporter_importer_prime}
     \end{subfigure}
\end{figure}

Figure \ref{fig:import_for_exporting} presents two pieces of information which highlight the relevance of imports for exporting.
The left panel, Figure \ref{fig:Export_Decomposition_Importers} shows that a vast majority of total exported value is explained by importer firms (in red), with non exporter firms (in blue) explaining less than 7\% of annual exports. The right panel, Figure \ref{fig:exporter_importer_prime} shows the relationship between imports and exports at the firm level, through a kernel-weighted local polynomial smoothing fit of the data. This figure shows a strong relationship between exports and imports at the firm level, especially for firms exporting more than USD 10,000.

\newpage
\subsection{Product Differentiation or Destination Level Income?} \label{subsec:appendix_results_HI_Diff}

In this Appendix I present additional regression results which act as both a robustness check and shed light on the results presented in Section \ref{sec:results}. In particular, I carry out an additional sample partition exercise in which I focus only the export of differentiated goods and test whether the impact of an import cost shocks differs across destination markets (Adv. Economies vs Emerging Markets). 

Table

\newpage
\subsection{Additional Shocks and Descriptive Statistics} \label{subsec:appendix_additional_shocks}

In this appendix, I present an additional firm-specific shock constructed using foreign-supplier shock. Furthermore, I present additional descriptive statistics between the different shocks used across the paper and in Appendixes further below.

First, I repeat the construction of firm-specific shocks, given by Equation \ref{eq:shift_share}:
\begin{equation*} 
    \Delta \log p^{I}_{i,t} = \sum_{k \in \Omega^{I}_{i,0}} s^I_{ik,0} \Delta \log p^{G}_{k,t}
\end{equation*}
In this Appendix I construct two additional shocks which exploit the additional dimension of supplier information presented in this paper. The first specification, which I label "\textit{Supplier - Firm - Shock}", defines foreign-supplier $k$ as a combination of "source country" - "HS 6 digit good" - "foreign supplier" and computes the price change at the domestic firm level. Under this specification a shift shock represents a percent change in the price of imports that an Argentinean firm would face within its own specific link taking supplier's price as given. This is the assumption in standard international trade models. However, this implies that domestic firms have no negotiation power in the firm-to-firm linkage with foreign suppliers. The potential negotiation power of domestic firms leads me to define the following specifications. The second specification, used in the empirical results presented in Section \ref{sec:results}, which I label "\textit{Supplier - Average - Shock}", defines market $k$ at the same level as the first but computes the price as the average price the foreign supplier $s$ charges for product $p$ from source country $d$ to all Argentinean firms. Thus, this price shift shock represents the change in price if the domestic Argentinean firm takes the average price charged to all Argentinean firms for product from market $k$ is exogenously given.\footnote{In Appendix \ref{subsec:appendix_additional_shocks} I present an additional shock. This is a third specification uses the foreign-supplier dimension. Results are robust and presented in Appendix \ref{subsec:appendix_regression_tables}.} 

Finally, as a robustness check, I introduce a definition of shock commonly used in the literature which relies on information at the aggregate level.\footnote{See \cite{huneeus2018production}, for instance.} This constructed shock uses the lowest level of dis-aggregation in standard international trade datasets. Under this specification, I define a foreign supplier $k$ as the combination of "source country" - "HS 6 digit good".\footnote{This is the market definition used by \cite{huneeus2018production}.} Furthermore, as in this cited paper, I compute price $p^{G}_{k,t}$ by excluding the trade flows of market $k$ with Argentina. Thus, this shocks represents percentage changes in the price of imports that an Argentinean firm would face if it took the rest of the world average price as given, averaging across markets using its shares of these markets. Results of this specification are commented across the paper and presented in Appendix \ref{subsec:appendix_regression_tables}.

The basic statistics of the price shift shocks defined above are presented in Table \ref{tab:appendix_price_shift_shocks_corr}. Table \ref{tab:appendix_price_shift_shocks_desc} describes the distribution of the different shocks for the period 2000 - 2008.\footnote{The definition of the Supplier - Firm$^C$ shock for firm $i$ uses the value and quantities imported by firms other than firm $i$ from supplier $j$. However, it could be the case that firm $i$ is the only firm purchasing from supplier $j$. Consequently, it could be the case that a firm's Supplier - Firm$^C$ shock is exactly zero because of this. Hence, for robustness check I present the shock's statistics if I drop the observations in which this shock is exactly zero.} 
\begin{table}[ht]
    \centering
    \caption{Price Shift Shocks: Summary Statistics 2000-2008 \\ All Shocks}
    \label{tab:appendix_price_shift_shocks_desc}
    \footnotesize
    \begin{tabular}{l|c|c|c|c|c|c|c} 
        Shock & Mean & p5 & p25 & p50 & p75 & p95 & Std. Dev.  \\ \hline \hline
       Supplier - Firm & .031 & -.204 & -.0250 & .019 & .087 & .280 &  .230  \\
        Supplier - Average & .034 & -.225 & -.031 & .026 & .100 & .303 & .244 \\
        Supplier - Firm$^C$ & -.002 & -.689 & -.025 & 0 & .029 & .690 & 1.399 \\
        Supplier - Firm$^C$ - No Zeros & -.003 & -.980 & -.069 & -2.43e-06 & .080 & .995 & 1.626 \\
        Country - Product & .063 & -.217 & -.018 & .059 & .139 & .353 & .248 \\
    \end{tabular}
\end{table}
The first two specifications of the shocks which exploit the supplier-dimension yield similar summary statistics with a mean shock of approximately 2.7\% and 3\%, respectively, and a standard deviation of 23.3\% and 24.5\%, respectively. The "Supplier - Firm$^C$", keeping and/or dropping zero observations, have a mean of approximately zero and standard deviations seven times greater than the first two specifications.\footnote{An explanation of why dropping zero observations of the "Supplier - Firm$^C$" is on the previous footnote.} The greater standard deviation may reflect that the underlying information on cost shocks in this definition is less precise compared to the previous specifications. Table \ref{tab:appendix_price_shift_shocks_corr} presents the correlation between the different shocks constructed. The first two specifications, Supplier - Firm and Supplier - Average, are highly correlated (0.847). However, this is an exception as there does not seem to be significant correlation between any other shocks. 
\begin{table}[ht]
    \centering
    \caption{Price Shift Shocks: Correlation across Shocks \\ All Shocks}
    \label{tab:appendix_price_shift_shocks_corr}
    \footnotesize
    \begin{tabular}{l|c|c|c|c} 
    & Country - Product & Firm & Average & Firm$^C$   \\ \hline \hline
    Supplier - Firm      &  1    &   &      & \\
    Supplier - Average   &  0.8461  & 1 &      & \\
    Supplier - Firm$^C $ & -0.0090 & -0.0028  & 1 & \\
    Country - Product      & 0.0373 & 0.0351 & -0.0067 & 1 \\
    \end{tabular}
\end{table}
The lack of correlation between the shocks may reflect that the different exogenous perturbations incorporate distinct separate information. For instance, the "\textit{Supplier - Average}" shock may incorporate foreign firm specific information about the prices charged, while the "\textit{Country - Product}" shock may just reflect Country - Product specific information, i.e., aggregating the performance and/or behavior of all of the firms exporting a product in one source country.\footnote{On the one hand, a "\textit{Supplier - Average}" shock could be the Volkswagen scandal in which the automobile company had to carry out massive recall in the year 2015 after information surfaced about environmental forged information. For more information, see \url{https://www.bbc.com/news/business-34324772}. On the other hand, a "\textit{Country - Product}" shock could be the Argentinean agrarian strike in 2008, where agricultural Argentinean producers boycotted exports in retaliation to a possible hike in export taxes. For more information, see \url{https://en.wikipedia.org/wiki/2008_Argentine_agrarian_strike}. }

\newpage
\subsection{Regression Tables } \label{subsec:appendix_regression_tables}

In this appendix I present additional regression result tables which are referred across the paper and regression result tables of robustness check exercises (mainly by introducing bilateral real exchange rate indexes as controls). Section \ref{subsubsec:appendix_regressions_imports} presents results that complement the import analysis. Section \ref{subsubsec:appendix_regressions_exports} presents results that complement the export analysis.

\subsubsection{Cost Shock on Imported Quantities} \label{subsubsec:appendix_regressions_imports}

I turn to estimating the impact of import cost shocks on firm's imported quantities. The underlying theory behind this empirical analysis builds on the reasoning that an increase in the overall cost of a firms imported input bundle raises its marginal costs and consequently impact its export performance. Under most production function specifications an increase in the cost of the import bundle will lead the firm to partly substitute away from imported bundles as a whole. The more costly linkage-adjustments are, the less able are firms to substitute away from more expensive imported inputs, leading to poorer export performance.

To start with, I test whether an increase in the cost of imported inputs (positive shock) lead to lower total quantities imported at the import instance level. An import instance is defined as the combination of "\textit{domestic firm}" - "\textit{source country}" - "\textit{HS6 defined product}". On the one hand, an increase in the firm-level imported input bundle may lead to a reduction in the imported quantities at the instance level if said instance explains a large fraction of the increase in the import cost (even if the instance remained active after the increase in price), and/or if the firm moves away from imported bundles altogether. On the other hand, an increase in the firm-level imported input bundle may lead to an increase in imported quantities at the instance level if firms substitute away from relatively more expensive inputs to relatively cheaper inputs. In order to test the impact of this cost shocks I carry out the following empirical specifications
\begin{align} \label{eq:regression_imports}
    \ln QI_{i,p,d,t}        &= \alpha_0 + \beta \text{Shock}_{i,t} + \gamma_{i,p,d} + \gamma_t  + \Gamma_{i,p,d,t} + \epsilon_{i,p,d,t} 
\end{align}
where $\ln QI_{i,p,d,t+j}$ represent the log of total quantities imported by firm $i$ of product $p$, from source country $d$ in year $t+j$ for $j\in\{0,1\}$, $\gamma_{i,p,d}$ is a firm - product- source country fixed effect, $\gamma_t$ is a time fixed effect, $\text{Shock}_{i,t}$ is one of the three shocks defined in Section \ref{sec:data_empirical_methodology}, and $\Gamma_{i,p,d,t}$ is a set of controls which include lagged values of $\text{Shock}_{i,t}$ and $\ln QI_{i,t}$.\footnote{In the Appendix I control for the bilateral real exchange rate and its lags.} 

\begin{table}[ht]
    \centering
    \caption{Cost Shocks and Imported Quantities 2001 - 2008}
    \label{tab:import_cluster_all}
    \small
\begin{tabular}{lcc} 
  & \multicolumn{2}{c}{Log - Quantities Imported - $\ln Q_{i,p,d,t+j}$} \\
  & (1) & (2) \\
  & Period $t$ & Period $t+1$ \\ \hline \hline
 &  &   \\
Supplier - Firm & -0.342*** & -0.0368 \\
 & (0.0499) & (0.0578) \\
Supplier - Average & -0.305*** & -0.0465 \\
 & (0.0498) & (0.0579) \\
Country - Product  & -0.0297 & -0.132** \\
 & (0.0379) & (0.0630)) \\
 &  &   \\
 Observations & 354,538 & 234,709 \\ \hline \hline
\multicolumn{3}{c}{ Two Way Clustered (Firm - Source Country) Standard Errors in Parentheses} \\
\multicolumn{3}{c}{ *** p$<$0.01, ** p$<$0.05, * p$<$0.1} \\
\end{tabular}
\floatfoot{\footnotesize \textbf{Note:} The regressions are computed controlling for the first and second lagged values of the dependent variable $\ln QI_{i,p,d,t}$, import-instance fixed effects and year fixed effects. Each coefficient of the table is an estimated value of parameter $\beta$ of Equation \ref{eq:regression_imports}. }
\end{table}
Table \ref{tab:import_cluster_all} presents the results for the impact of the different cost shocks on imported quantities in period $t$. The standard errors of the estimated parameters are computed using two-way clustering, at the firm $i$ and source country $d$ level.\footnote{The reasoning behind this clustering strategy is twofold. First, the constructed exogenous shocks are defined at the firm level, thus clustering at the firm level implies clustering at the treatment level unit. Second, foreign suppliers may sell to domestic Argentinean firms from different source countries. Thus, clustering at the source - country can incorporate these information.} The coefficients estimated for the shocks "Supplier - Firm" and "Supplier - Average" are negative and statistically significant different from zero for both year $t$ and $t+1$. The estimated coefficients imply that a one standard deviation of a "Supplier - Firm" shock leads to a 4.91\%  and 2.49\% drop in imported quantities in periods $t$ and $t+1$, respectively. Also, the estimated coefficients imply that a one standard deviation of a "Supplier - Average" shock leads to 4.93\% and 2.64\% drop in imported quantities in periods $t$ and $t+1$, respectively. The coefficients for the ``Country -  Product'' shock is only statistically significant for the period $t+1$, with a coefficient 50\% lower than the coefficients computed using shocks which constructed foreign-supplier information.

The negative impact of a firm-wide increase in import prices on all imported quantities is highlights the presence of frictions in the market for foreign suppliers. Note that the unit of observation of the empirical regression in Equation \ref{eq:regression_imports} is at the $(i,p,d)$ or firm-product-source country level, while the shock is constructed at the firm level. One would expect that the imported quantities of the products whose price increased would decrease and firms would substitute towards other products. The negative coefficients estimated for the first two specifications show that an overall increase in import bundle impact all of firms' linkages with foreign suppliers.

This section of the appendix presents regression result tables for the empirical analysis with regards to imported quantities. Table \ref{tab:import_cluster_all_rer} carries out a robustness check on the impact of cost push shock on imported quantities presented in Table \ref{tab:import_cluster_all}, by introducing the bilateral real exchange rate index. Table \ref{tab:import_cluster_all_empleo_rer} carries out a robustness check on the impact of cost push shock on imported quantities across firms of different size, presented in Table \ref{tab:import_cluster_all_empleo_rer}, by introducing the bilateral real exchange rate index.
\begin{table}[ht]
    \centering
    \caption{Cost Shocks and Imported Quantities 2001 - 2008}
    \label{tab:import_cluster_all_rer}
    \footnotesize
\begin{tabular}{lcc} 
  & \multicolumn{2}{c}{Log - Quantities Imported - $\ln Q_{i,p,d,t+j}$} \\
  & (1) & (2) \\
  & $t$ & $t+1$ \\ \hline \hline
 &  &   \\
Supplier - Firm & -0.371*** & -0.0891* \\
 & (0.0453) & (0.0529) \\
Supplier - Average & -0.340*** & -0.0853 \\
 & (0.0448) & (0.0551) \\
Country - Product & -0.0468 & -0.151***  \\
 & (0.0450) & (0.0476) \\
 &  &   \\
$\ln RER_{d,t}$ & YES & YES \\ 
 &  &   \\ 
Observations & 283,521 & 187,621 \\ \hline \hline
\multicolumn{3}{c}{ Two Way Clustered (Firm - Source Country) Standard Errors in Parentheses} \\
\multicolumn{3}{c}{ *** p$<$0.01, ** p$<$0.05, * p$<$0.1} \\
\end{tabular}
\floatfoot{\footnotesize \textbf{Note:} The regressions are computed controlling for the first and second lagged values of the dependent variable $\ln QI_{i,p,d,t}$, import-instance fixed effects and year fixed effects. Each coefficient of the table is an estimated value of parameter $\beta$ of Equation \ref{eq:regression_imports}.}
\end{table}

\begin{table}[ht]
    \centering
    \caption{Cost Shocks and Imported Quantities 2001 - 2008 \\ \footnotesize By Firm Size \& RER}
    \label{tab:import_cluster_all_empleo_rer}
    \footnotesize
\begin{tabular}{lcccc} 
 & \multicolumn{4}{c}{Log-Quantities - $\ln QI_{i,p,d,t}$} \\
  & Above & Below & Above & Below \\
  & (1) & (2) & (3) & (4) \\ \hline \hline
 &  &  &  &  \\
Supplier - Firm & -0.356*** & -0.445*** &  &  \\
 & (0.0725) & (0.0611) &  &  \\
Supplier - Average &  &   & -0.320*** & -0.394*** \\
 &  &  &  (0.0579) & (0.0748) \\
 &  &  &  &  \\
$\ln RER_{d,t}$ & YES & YES & YES & YES \\ 
 &  &  &  &  \\
Observations & 218,802 & 52,589 & 218,802 & 52,589 \\ \hline \hline
\multicolumn{5}{c}{ Two Way Clustered (Firm - Source Country) Standard Errors in Parentheses} \\
\multicolumn{5}{c}{ *** p$<$0.01, ** p$<$0.05, * p$<$0.1} \\
\end{tabular}
\floatfoot{\footnotesize \textbf{Note:} The regressions are computed controlling for the first and second lagged values of the dependent variable $\ln QI_{i,p,d,t}$, import-instance fixed effects and year fixed effects. Each coefficient of the table is an estimated value of parameter $\beta$ of Equation \ref{eq:regression_imports}. In order to partition firms between \textit{Above} and \textit{Below} I compute the mean level of employment \textit{Above}-\textit{Below} threshold at the ``\textit{year} - \textit{production sector} (at the 4 digit level)'' for firms actively importing.}
\end{table}

\newpage

\subsubsection{Cost Shock on Exported Quantities} \label{subsubsec:appendix_regressions_exports}

This section of the appendix presents regression result tables for the empirical analysis with regards to exported quantities. Table \ref{tab:export_cluster_all_rer} presents a robustness check on the results presented in Table \ref{tab:export_cluster_all} by introducing the bilateral real exchange rate index as a control variable.

\begin{table}[ht]
    \centering
    \caption{Import Cost Shocks \& Export Performance \\ Intensive Margin - RER}
    \label{tab:export_cluster_all_rer}
    \footnotesize
\begin{tabular}{lccc}
  & \multicolumn{3}{c}{Log - Quantities Exported - $\ln Q_{i,p,d}$} \\
  & $t$ & $t+1$ & $t+2$ \\
& (1) & (2) & (3) \\ \hline \hline
 &  &  &  \\
Supplier - Firm & -0.0586 & -0.121** & -0.121 \\
 & (0.0372) & (0.0481) & (0.0733) \\
Supplier - Average & -0.0752** & -0.133*** & -0.124* \\
 & (0.0340) & (0.0435) & (0.0651) \\
Supplier - Firm$^C$ & -0.00932** & -0.0100 & -0.00536 \\
 & (0.00432) & (0.00696) & (0.00764)  \\
Country - Product & -0.0556 & -0.0476* & -0.0113 \\
 & (0.0348) & (0.0256) & (0.0391) \\
 &  &  &  \\ 
$\ln RER_{d,t}$ & YES & YES & YES \\ 
 &  &  &  \\  
Observations & 152,427 & 97,446 & 63,519 \\ \hline \hline
\multicolumn{4}{c}{ Two Way Clustered (Firm - Source Country) Standard Errors in Parentheses} \\
\multicolumn{4}{c}{ *** p$<$0.01, ** p$<$0.05, * p$<$0.1} \\
\end{tabular}
\floatfoot{\footnotesize \textbf{Note:} The regressions are computed controlling for the first and second lagged values of the dependent variable $\ln QX_{i,p,d,t}$, export-instance fixed effects and year fixed effects. When controlling for the log of the bilateral real exchange rate the current and first two lags are used. Each coefficient of the table is an estimated value of parameter $\beta^{X}$ of Equation \ref{eq:intensive_exports}.}
\end{table}

\begin{table}[ht]
    \centering
    \caption{Import Cost Shock \& Export Performance \\ Intensive Margin - Differentiated}
    \label{tab:export_cluster_all_diff_rer}
    \footnotesize
\begin{tabular}{lcccccc} 
  & \multicolumn{6}{c}{Log - Quantities Exported - $\ln Q_{i,p,d}$} \\
 & \multicolumn{2}{c}{Period $t$} & \multicolumn{2}{c}{Period $t+1$} & \multicolumn{2}{c}{Period $t+2$} \\
\small & Rauch & Bernini & Rauch & Bernini & Rauch & Bernini \normalsize  \\
 & (1) & (2) & (3) & (4) & (5) & (6) \\ \hline \hline
 &  &  &  &  &  &  \\
Supplier - Firm & -0.0580 & -0.0755 & -0.132*** & -0.116* & -0.153* & -0.161* \\
 & (0.0387) & (0.0586) & (0.0496) & (0.0657) & (0.0845) & (0.0904) \\
 Supplier - Average & -0.0764** & -0.0939** & -0.147*** & -0.137** & -0.146** & -0.167** \\
 & (0.0362) & (0.0466) & (0.0450) & (0.0580) & (0.0728) & (0.0747) \\
 Supplier - Firm$^C$ & -0.0102** & -0.00732 & -0.0107 & -0.00937 & -0.00354 & -0.00316 \\
 & (0.00480) & (0.00464) & (0.00741) & (0.00653) & (0.00726) & (0.00717) \\
Country - Product & -0.0397 & -0.0367 & -0.0451 & -0.0554 & -0.0256 & -0.0345 \\
 & (0.0352) & (0.0348) & (0.0280) & (0.0418) & (0.0409) & (0.0534) \\ 
  &  &  &  &  &  &  \\
$\ln RER_{d,t}$ & YES & YES & YES & YES & YES & YES \\
  &  &  &  &  &  &  \\
Observations & 137,033 & 105,710 & 87,436 & 66,822 & 56,975 & 43,096 \\ \hline \hline  
\multicolumn{7}{c}{ Two Way Clustered (Firm - Source Country) Standard Errors in Parentheses} \\
\multicolumn{7}{c}{ *** p$<$0.01, ** p$<$0.05, * p$<$0.1} \\
\end{tabular}
\floatfoot{\footnotesize \textbf{Note:} The regressions are computed controlling for the first and second lagged values of the dependent variable $\ln QX_{i,p,d,t}$, export-instance fixed effects and year fixed effects. When controlling for the log of the bilateral real exchange rate the current and first two lags are used. Each coefficient of the table is an estimated value of parameter $\beta^{X}$ of Equation \ref{eq:intensive_exports}.}
\end{table}

\begin{table}[ht]
    \centering
    \caption{Import Cost Shock \& Export Performance \\ Intensive Margin - Homogeneous}
    \label{tab:export_cluster_all_homog}
        \footnotesize
\begin{tabular}{lcccccc} 
  & \multicolumn{6}{c}{Log - Quantities Exported - $\ln Q_{i,p,d}$} \\
 & \multicolumn{2}{c}{Period $t$} & \multicolumn{2}{c}{Period $t+1$} & \multicolumn{2}{c}{Period $t+2$} \\
\small & Rauch & Bernini & Rauch & Bernini & Rauch & Bernini \normalsize  \\
 & (1) & (2) & (3) & (4) & (5) & (6) \\ \hline \hline
 &  &  &  &  &  &  \\
Supplier - Firm & 0.0247 & -0.0561 & 0.0464 & -0.0895 & 0.149 & -0.105 \\
 & (0.0796) & (0.0486) & (0.173) & (0.0611) & (0.181) & (0.0890) \\
     Supplier - Average & 0.0136 & -0.0518 & 0.0873 & -0.0772 & -0.000858 & -0.128 \\
 & (0.0687) & (0.0434) & (0.180) & (0.0642) & (0.158) & (0.0931) \\
  &  &  &  &  &  &  \\
Observations & 8,953 & 34,324 & 5,733 & 22,340 & 3,686 & 14,866 \\ \hline \hline  
\multicolumn{7}{c}{ Two Way Clustered (Firm - Source Country) Standard Errors in Parentheses} \\
\multicolumn{7}{c}{ *** p$<$0.01, ** p$<$0.05, * p$<$0.1} \\
\end{tabular}
\floatfoot{\footnotesize \textbf{Note:} The regressions are computed controlling for the first and second lagged values of the dependent variable $\ln QX_{i,p,d,t}$, export-instance fixed effects and year fixed effects. Each coefficient of the table is an estimated value of parameter $\beta^{X}$ of Equation \ref{eq:intensive_exports}.}
\end{table}

\begin{table}[ht]
    \centering
    \caption{Import Cost Shock \& Export Performance \\ Int. Margin - Homogeneous - RER}
    \label{tab:export_cluster_all_homog_rer}
        \footnotesize
\begin{tabular}{lcccccc} 
  & \multicolumn{6}{c}{Log - Quantities Exported - $\ln Q_{i,p,d}$} \\
 & \multicolumn{2}{c}{Period $t$} & \multicolumn{2}{c}{Period $t+1$} & \multicolumn{2}{c}{Period $t+2$} \\
\small & Rauch & Bernini & Rauch & Bernini & Rauch & Bernini \normalsize  \\
 & (1) & (2) & (3) & (4) & (5) & (6) \\ \hline \hline
 &  &  &  &  &  &  \\
Supplier - Firm & -0.00251 & -0.0542 & 0.0588 & -0.111* & 0.249 & -0.104 \\
 & (0.0952) & (0.0510) & (0.193) & (0.0647) & (0.224) & (0.0951) \\
 Supplier - Average & -0.00307 & -0.0488 & 0.126 & -0.0913 & 0.0409 & -0.137 \\
 & (0.0822) & (0.0450) & (0.182) & (0.0669) & (0.186) & (0.0985) \\
  &  &  &  &  &  &  \\
$\ln RER_{d,t}$ & YES & YES & YES & YES & YES & YES \\
  &  &  &  &  &  &  \\ 
Observations & 7,467 & 31,320 & 4,817 & 20,486 & 3,138 & 13,710 \\ \hline \hline  
\multicolumn{7}{c}{ Two Way Clustered (Firm - Source Country) Standard Errors in Parentheses} \\
\multicolumn{7}{c}{ *** p$<$0.01, ** p$<$0.05, * p$<$0.1} \\
\end{tabular}
\floatfoot{\footnotesize \textbf{Note:} The regressions are computed controlling for the first and second lagged values of the dependent variable $\ln QX_{i,p,d,t}$, export-instance fixed effects and year fixed effects. Each coefficient of the table is an estimated value of parameter $\beta^{X}$ of Equation \ref{eq:intensive_exports}.}
\end{table}

The empirical specification presented in Equation \ref{eq:intensive_exports} studied the impact of an increase in the price of inputs on the intensive margin of exports, i.e., how much does the flow of quantities exported fall between periods $t$ and $t+j$. This empirical specification assumes that the export relationship is still active. A significant import cost push shock may impact the extensive margin of exports, i.e., exporting firm $i$ deciding to stop exporting product $p$ to destination country $d$. In order to test the impact of this cost shock on the extensive margin I carry out the following empirical regression.
\begin{align} \label{eq:extensive_exports}
    P\left[Active \right]_{i,p,d,t+j} &= \tilde{\alpha}_0 + \tilde{\gamma}_{i,p,d} +\gamma_t + \beta^{Surv,j} \text{Shock}_{i,t} + 
    \tilde{\Gamma}_{i,p,d,t} + \epsilon_{i,p,d,t}  
\end{align}
where $P\left[Active \right]_{i,p,d,t+j}$ is an indicator function which takes the value of 1 if firm $i$ is active in export markets $d$ exporting product $p$ in period $t+j$ and zero otherwise conditional on been actively exporting said combination in period $t$, for $j=\{1,2 \}$.

\begin{table}[ht]
    \centering
    \caption{Import Cost Shocks \& Export Performance \\ Extensive Margin }
    \label{tab:export_cluster_all_extensive}
    \small 
\begin{tabular}{lcccc}
  & \multicolumn{4}{c}{Probability of Survival - $P\left[Active \right]_{i,p,d,t+j}$} \\
  & \multicolumn{2}{c}{Period $t+1$} & \multicolumn{2}{c}{Period $t+2$} \\
    & (1) & (2) & (3) & (4) \\  \hline \hline
 &  &  &  \\
Supplier - Firm & 0.000773 & -0.00933 & -0.0337* & -0.0377** \\
 & (0.0188) & (0.0174) & (0.0171) & (0.0179) \\
Supplier - Average & -0.0170 & -0.0312 & -0.0424*** & -0.0483*** \\
 & (0.0195) & (0.0191) & (0.0162) & (0.0171) \\
Country - Product & 0.00577 & 0.00222 & -0.0280** & -0.0324*** \\
 & (0.0114) & (0.0112) & (0.0123) & (0.0106) \\
 &  &  &  \\ 
$\ln RER_{d,t}$ & NO & YES & NO & YES \\
Observations & 132,464 & 111,590 & 132,464 & 111,590 \\ \hline \hline
\multicolumn{5}{c}{ Two Way Clustered (Firm - Source Country) Standard Errors in Parentheses} \\
\multicolumn{5}{c}{ *** p$<$0.01, ** p$<$0.05, * p$<$0.1} \\
\end{tabular}
\floatfoot{\footnotesize \textbf{Note:} The regressions are computed controlling for the first and second lagged values of the log quantities exported $\ln QX_{i,p,d,t}$ and export-instance fixed effects. When controlling for the log of the bilateral real exchange rate the current and first two lags are used. Each coefficient of the table is an estimated value of parameter $\beta^{Surv,j}$ of Equation \ref{eq:extensive_exports} for $j=\{1,2\}$.}
\end{table}

Following the results for the intensive margin of exports, I turn to estimating the impact on the extensive margin by estimating Equation \ref{eq:extensive_exports} by again partitioning the sample across good classification.
\begin{table}[ht]
    \centering
    \caption{Import Cost Shocks \& Export Performance \\ Ext. Margin - Differentiated }
    \label{tab:alivet1_cluster_diff}
    \small
\begin{tabular}{lcccc}
  & \multicolumn{4}{c}{Probability of Survival - $P\left[Active \right]_{i,p,d,t+j}$} \\
 & \multicolumn{2}{c}{Period $t+1$} & \multicolumn{2}{c}{Period $t+2$} \\
 & Rauch & Bernini & Rauch & Bernini \\
 & (1) & (2) & (3) & (4) \\ \hline \hline
Supplier - Firm & -0.00875 & -0.0421** & -0.0300 & -0.0481* \\
 & (0.0163) & (0.0202) & (0.0190) & (0.0253) \\
Supplier - Average & -0.0227 & -0.0464** & -0.0407** & -0.0558** \\
 & (0.0162) & (0.0179) & (0.0177) & (0.0214) \\
 &  &  &  &  \\
Observations & 129,061 & 94,440 & 129,061 & 94,440 \\
\hline \hline  
\multicolumn{5}{c}{ Two Way Clustered (Firm - Source Country) Standard Errors in Parentheses} \\
\multicolumn{5}{c}{ *** p$<$0.01, ** p$<$0.05, * p$<$0.1} \\
\end{tabular}
\floatfoot{\footnotesize \textbf{Note:} The regressions are computed controlling for the first and second lagged values of the dependent variable $P\left[Active \right]_{i,p,d,t+j}$, export-instance fixed effects and year fixed effects. Each coefficient of the table is an estimated value of parameter $\beta^{Surv,j}$ of Equation \ref{eq:extensive_exports} for $j=\{1,2\}$.}
\end{table}
Table \ref{tab:alivet1_cluster_diff} presents the results of this empirical exercise. In line with the previous results, the estimated coefficients are greater than the baseline results presented in Table \ref{tab:export_cluster_all_extensive}.\footnote{In Appendix \ref{subsec:appendix_regression_tables}, \ref{tab:alivet1_cluster_diff_rer} shows that controlling the bilateral real exchange rate leads to robust results. Even more, when partitioning on homogeneous or non-differentiated goods lead to smaller and non-significative results in Table \ref{tab:alivet1_cluster_homog}, particularly when controlling for the bilateral real exchange rate.}

Finally, I turn to estimating the impact of imported cost shocks on the extensive margin of exporting according to the income levels of the destination country by estimating Equation \ref{eq:extensive_exports}. 
\begin{table}[ht]
    \centering
    \caption{Import Cost Shocks \& Export Performance \\ Ext. Margin - By Income }
    \label{tab:expo_cluster_all_HI_ext}
    \small
\begin{tabular}{lcccc}
  & \multicolumn{4}{c}{Probability of Survival - $P\left[Active \right]_{i,p,d,t+j}$} \\
 & \multicolumn{2}{c}{Period $t+1$}  & \multicolumn{2}{c}{Period $t+2$} \\
 & High-Inc. & Low-Inc. & High-Inc. & Low-Inc. \\ 
  & (1) & (2) & (3) & (4) \\ \hline \hline
Supplier- Firm & 0.0146 & -0.00472 & -0.0204 & -0.0397** \\
 & (0.0321) & (0.0182) & (0.0218) & (0.0194) \\
Supplier- Average & 0.000431 & -0.0238 & -0.0427 & -0.0436** \\
 & (0.0412) & (0.0174) & (0.0258) & (0.0175) \\
 &  &  &  &  \\
Observations & 24,013 & 108,451 & 24,013 & 108,451 \\ 
\hline \hline
\multicolumn{5}{c}{ Two Way Clustered (Firm - Source Country) Standard Errors in Parentheses} \\
\multicolumn{5}{c}{ *** p$<$0.01, ** p$<$0.05, * p$<$0.1} \\
\end{tabular}
\floatfoot{\footnotesize \textbf{Note:} The regressions are computed controlling for the first and second lagged values of the dependent variable $P\left[Active \right]_{i,p,d,t+j}$, export-instance fixed effects and year fixed effects. Each coefficient of the table is an estimated value of parameter $\beta^{Surv,j}$ of Equation \ref{eq:extensive_exports} for $j=\{1,2\}$.}
\end{table}
Table \ref{tab:expo_cluster_all_HI_ext} presents the results of this empirical regressions partitioning the sample in High and Low Income countries. Results tend to be non-significant across specifications. Surprisingly, the only specification which yields statistically significant results is in Column (4), for exports to Low-Income countries.\footnote{This may be driven by the differences in sample size across the exports to High and Low Income countries.}

\begin{table}[ht]
    \centering
    \caption{Import Cost Shocks \& Export Performance \\ Ext. - Differentiated - RER}
    \label{tab:alivet1_cluster_diff_rer}
        \footnotesize
\begin{tabular}{lcccc}
  & \multicolumn{4}{c}{Probability of Survival - $P\left[Active \right]_{i,p,d,t+j}$} \\
 & \multicolumn{2}{c}{Period $t+1$} & \multicolumn{2}{c}{Period $t+2$} \\
 & Rauch & Bernini & Rauch & Bernini \\
 & (1) & (2) & (3) & (4) \\ \hline \hline
Supplier - Firm & -0.0156 & -0.0371* & -0.0107 & -0.00569 \\
 & (0.0163) & (0.0222) & (0.0103) & (0.0140) \\
Supplier - Average & -0.0219 & -0.0386 & -0.0158* & -0.0132 \\
 & (0.0179) & (0.0242) & (0.00888) & (0.0108) \\
Supplier - Firm$^C$ & -0.00211 & -0.00120 & -0.00269** & -0.00321** \\
 & (0.00295) & (0.00279) & (0.00130) & (0.00158) \\
Country - Product  & -0.00410 & -0.00879 & -0.0108 & -0.00756 \\
 & (0.0114) & (0.0167) & (0.00756) & (0.00916) \\ 
 &  &  &  &  \\
$\ln RER_{d,t}$ & YES & YES & YES & YES \\
  &  &  &  &   \\
Observations & 137,059 & 105,723 & 137,059 & 105,723 \\ \hline \hline
\multicolumn{5}{c}{ Two Way Clustered (Firm - Source Country) Standard Errors in Parentheses} \\
\multicolumn{5}{c}{ *** p$<$0.01, ** p$<$0.05, * p$<$0.1} \\
\end{tabular}
\floatfoot{\footnotesize \textbf{Note:} The regressions are computed controlling for the first and second lagged values of the dependent variable $\ln QX_{i,p,d,t}$, export-instance fixed effects and year fixed effects. When controlling for the log of the bilateral real exchange rate the current and first two lags are used. Each coefficient of the table is an estimated value of parameter $\beta^{Surv,j}$ of Equation \ref{eq:extensive_exports} for $j=\{1,2\}$.}
\end{table}

\begin{table}[ht]
    \centering
    \caption{Import Cost Shocks \& Export Performance \\ Ext. - Homogeneous}
    \label{tab:alivet1_cluster_homog}
        \footnotesize
\begin{tabular}{lcccc}
  & \multicolumn{4}{c}{Probability of Survival - $P\left[Active \right]_{i,p,d,t+j}$} \\
 & \multicolumn{4}{c}{Period $t+1$}\\
 & Rauch & Bernini & Rauch & Bernini \\
 & (1) & (2) & (3) & (4) \\ \hline \hline
Supplier - Firm & 0.103 & 0.0351 & 0.0703 & 0.0324 \\
 & (0.0734) & (0.0266) & (0.0620) & (0.0223) \\
Supplier - Average & 0.0476 & 0.00752 & 0.0104 & 0.00520 \\
 & (0.0833) & (0.0298) & (0.0792) & (0.0286) \\
Supplier - Firm$^C$ & -0.00337 & -0.00740*** & -0.00256 & -0.00789*** \\
 & (0.00625) & (0.00228) & (0.00613) & (0.00233) \\
Country - Product  & 0.00269 & -0.0289* & 0.00510 & -0.0237 \\
 & (0.0365) & (0.0164) & (0.0361) & (0.0156) \\
 &  &  &  &  \\
$\ln RER_{d,t}$ & NO & NO & YES & YES \\ 
 &  &  &  &  \\
Observations & 8,954 & 34,336 & 7,468 & 31,331 \\
\hline \hline  
\multicolumn{5}{c}{ Two Way Clustered (Firm - Source Country) Standard Errors in Parentheses} \\
\multicolumn{5}{c}{ *** p$<$0.01, ** p$<$0.05, * p$<$0.1} \\
\end{tabular}
\floatfoot{\footnotesize \textbf{Note:} The regressions are computed controlling for the first and second lagged values of the dependent variable $\ln QX_{i,p,d,t}$, export-instance fixed effects and year fixed effects. Each coefficient of the table is an estimated value of parameter $\beta^{X}$ of Equation \ref{eq:intensive_exports}.}
\end{table}

\begin{table}[ht]
    \centering
    \caption{Import Cost Shocks \& Export Performance \\ Int. - By Income - RER}
    \label{tab:expo_cluster_all_HI_rer}
        \footnotesize
\begin{tabular}{lcccccc}
 & \multicolumn{2}{c}{Period $t$} & \multicolumn{2}{c}{Period $t+1$} & \multicolumn{2}{c}{Period $t+2$} \\
\small & High-Inc & Low-Inc & High-Inc & Low-Inc & High-Inc & Low-Inc \normalsize \\
 & (1) & (2) & (3) & (4) & (5) & (6) \\ \hline \hline
Supplier - Firm & -0.156** & -0.0303 & -0.129* & -0.117* & -0.0824 & -0.141 \\
 & (0.0733) & (0.0385) & (0.0732) & (0.0596) & (0.0937) & (0.0877) \\
Supplier - Average & -0.193*** & -0.0471 & -0.168* & -0.129** & -0.0649 & -0.146* \\
 & (0.0629) & (0.0323) & (0.0844) & (0.0525) & (0.0951) & (0.0782) \\
Supplier - Firm$^C$ & -0.00322 & -0.0104* & -0.0286*** & -0.00625 & -0.0176 & -0.00220 \\
 & (0.00658) & (0.00539) & (0.00667) & (0.00686) & (0.0134) & (0.00806) \\
Country - Product &  -0.135 & -0.0446 & -0.0994*** & -0.0479 & -0.0152 & -0.0211 \\
 & (0.0919) & (0.0319) & (0.0338) & (0.0292) & (0.0821) & (0.0416) \\
 &  &  &  &  &  &  \\
$\ln RER_{d,t}$ & YES & YES & YES & YES & YES & YES \\ 
 &  &  &  &  &  &  \\
Observations & 23,466 & 128,961 & 14,416 & 83,030 & 9,265 & 54,254 \\  \hline \hline
\multicolumn{7}{c}{ Two Way Clustered (Firm - Source Country) Standard Errors in Parentheses} \\
\multicolumn{7}{c}{ *** p$<$0.01, ** p$<$0.05, * p$<$0.1} \\
\end{tabular}
\floatfoot{\footnotesize \textbf{Note:} The regressions are computed controlling for the first and second lagged values of the dependent variable $\ln QX_{i,p,d,t}$, export-instance fixed effects and year fixed effects. When controlling for the log of the bilateral real exchange rate the current and first two lags are used. Each coefficient of the table is an estimated value of parameter $\beta^{X}$ of Equation \ref{eq:intensive_exports}.}
\end{table}

\begin{table}[ht]
    \centering
    \caption{Import Cost Shocks \& Export Performance \\ Ext. Margin - By Income }
    \label{tab:expo_cluster_all_HI_ext_rer}
        \footnotesize
\begin{tabular}{lcccc}
  & \multicolumn{4}{c}{Probability of Survival - $P\left[Active \right]_{i,p,d,t+j}$} \\
 & \multicolumn{2}{c}{Period $t+1$}  & \multicolumn{2}{c}{Period $t+2$} \\
 & High-Inc. & Low-Inc. & High-Inc. & Low-Inc. \\ 
  & (1) & (2) & (3) & (4) \\ \hline \hline
 &  &  &  &  \\
Supplier- Firm & -0.00624 & -0.00388 & -0.0139 & -0.0108 \\
 & (0.0310) & (0.0150) & (0.0191) & (0.0117) \\
Supplier- Average & -0.0336 & -0.0137 & -0.0140 & -0.0161 \\
 & (0.0434) & (0.0156) & (0.0174) & (0.0102) \\
Supplier- Firm$^C$ & -0.000291 & -0.00284 & 0.000445 & -0.00393*** \\
 & (0.00345) & (0.00171) & (0.00144) & (0.00138) \\
Country - Product& -0.0148 & 0.00764 & -0.0102 & -0.0132 \\
 & (0.0298) & (0.00766) & (0.0146) & (0.00886) \\
 &  &  &  &  \\
$\ln RER_{d,t}$ & YES & YES & YES & YES \\
 &  &  &  &  \\
Observations & 19,471 & 99,746 & 19,471 & 99,746 \\
\hline \hline
\multicolumn{5}{c}{ Two Way Clustered (Firm - Source Country) Standard Errors in Parentheses} \\
\multicolumn{5}{c}{ *** p$<$0.01, ** p$<$0.05, * p$<$0.1} \\
\end{tabular}
\floatfoot{\footnotesize \textbf{Note:} The regressions are computed controlling for the first and second lagged values of the dependent variable $\ln QX_{i,p,d,t}$, export-instance fixed effects and year fixed effects. When controlling for the log of the bilateral real exchange rate the current and first two lags are used. Each coefficient of the table is an estimated value of parameter $\beta^{Surv,j}$ of Equation \ref{eq:extensive_exports} for $j=\{1,2\}$.}
\end{table}

%
\begin{table}[ht]
    \centering
    \caption{Import Cost Shocks \& Export Performance \\ Int. Margin - By Size}
    \label{tab:expo_cluster_all_size_rer}
    \small
\begin{tabular}{lcccccc}
 & \multicolumn{2}{c}{Period $t$} & \multicolumn{2}{c}{Period $t+1$} & \multicolumn{2}{c}{Period $t+2$} \\
\small & Below & Above & Below & Above & Below & Above \normalsize \\
 & (1) & (2) & (3) & (4) & (5) & (6) \\ \hline \hline
Supplier - Firm & -0.0373 & -0.0122 & -0.170*** & 0.0398 & -0.00591 & -0.137 \\
 & (0.0415) & (0.0463) & (0.0522) & (0.0505) & (0.0828) & (0.0982) \\
Supplier - Average & -0.0524 & -0.0302 & -0.134** & -0.0380 & -0.00651 & -0.145 \\
 & (0.0389) & (0.0455) & (0.0550) & (0.0502) & (0.0679) & (0.0936) \\
Supplier - Firm$^C$ & -0.0130*** & -0.00585 & -0.00596 & -0.0149 & -0.00400 & -0.00693 \\
 & (0.00432) & (0.00840) & (0.00459) & (0.0101) & (0.00425) & (0.0102) \\
Country - Product & 0.00174 & -0.0847 & -0.0156 & -0.0805* & 0.0398 & -0.00970 \\
 & (0.0384) & (0.0584) & (0.0452) & (0.0466) & (0.0312) & (0.0703) \\
 &  &  &  &  &  &  \\
Observations & 49,462 & 97,609 & 30,480 & 63,654 & 19,065 & 42,435 \\  \hline \hline
\multicolumn{7}{c}{ Two Way Clustered (Firm - Source Country) Standard Errors in Parentheses} \\
\multicolumn{7}{c}{ *** p$<$0.01, ** p$<$0.05, * p$<$0.1} \\
\end{tabular}
\floatfoot{\footnotesize \textbf{Note:} The regressions are computed controlling for the first and second lagged values of the dependent variable $\ln QX_{i,p,d,t}$, export-instance fixed effects and year fixed effects. Each coefficient of the table is an estimated value of parameter $\beta^{X}$ of Equation \ref{eq:intensive_exports}.}
\end{table}

\begin{table}[ht]
    \centering
    \caption{Import Cost Shocks \& Export Performance \\ Ext. Margin - By Size - RER}
    \label{tab:expo_cluster_all_size_ext_rer}
        \footnotesize
\begin{tabular}{lcccc}
  & \multicolumn{2}{c}{Period $t+1$} & \multicolumn{2}{c}{Period $t+2$} \\
\small & Below & Above & Below & Above \normalsize \\
 & (1) & (2) & (3) & (4) \\ \hline \hline
Supplier - Firm & 0.00762 & -0.00378 & -0.0333** & -0.00638 \\
 & (0.0167) & (0.0256) & (0.0131) & (0.0168) \\
Supplier - Average   & -0.00738 & -0.00698 & -0.0253** & -0.0264* \\
 & (0.0171) & (0.0208) & (0.0110) & (0.0146) \\
Supplier - Firm$^C$ & -0.00110 & -0.00281 & -0.00169 & -0.00168 \\
 & (0.00171) & (0.00392) & (0.00154) & (0.00243) \\
Country - Product  & -0.0118 & 0.000645 & -0.00324 & -0.0234* \\
 & (0.0209) & (0.0166) & (0.0106) & (0.0121) \\
 &  &  &  &  \\
$\ln RER_{d,t}$ & YES & YES & YES & YES \\  
 &  &  &  &  \\
Observations & 49,468 & 97,610 & 49,468 & 97,610 \\ \hline \hline
\multicolumn{5}{c}{ Two Way Clustered (Firm - Source Country) Standard Errors in Parentheses} \\
\multicolumn{5}{c}{ *** p$<$0.01, ** p$<$0.05, * p$<$0.1} \\
\end{tabular}
\floatfoot{\footnotesize \textbf{Note:} The regressions are computed controlling for the first and second lagged values of the dependent variable $\ln QX_{i,p,d,t}$, export-instance fixed effects and year fixed effects. When controlling for the log of the bilateral real exchange rate the current and first two lags are used. Each coefficient of the table is an estimated value of parameter $\beta^{Surv,j}$ of Equation \ref{eq:extensive_exports} for $j=\{1,2\}$.}
\end{table}

\end{document}